\documentclass[aps,reprint,superscriptaddress,longbibliography,bibnotes]{revtex4-2}

\usepackage{graphicx}
\usepackage{dcolumn}
\usepackage{bm, amssymb, color}
\usepackage{hyperref}
\usepackage{natbib}
\usepackage[mathscr]{euscript}
\usepackage{booktabs,tabulary}
\usepackage{multirow}
\usepackage{amsmath,soul}
\usepackage[export]{adjustbox}

\usepackage[caption=false]{subfig}
\usepackage{array}
\usepackage{cases}

\usepackage{float}

\newcommand{\fn}[2]{\mathinner{#1\mathopen{\left(#2\right)}}}

\newcommand{\spD}[1]{\fn{\tilde{\chi}_{_V}}{#1}}

\usepackage[colorinlistoftodos,shadow,textwidth=18mm]{todonotes}

\newcolumntype{L}[1]{>{\raggedright\let\newline\\\arraybackslash\hspace{0pt}}m{#1}}
\newcolumntype{C}[1]{>{\centering\let\newline\\\arraybackslash\hspace{0pt}}m{#1}}
\newcolumntype{R}[1]{>{\raggedleft\let\newline\\\arraybackslash\hspace{0pt}}m{#1}}
\setstcolor{green}

\graphicspath{{./figs/}}

\begin{document}
\title{Hyperuniformity of maximally random jammed packings of hyperspheres across spatial dimensions}

\author{Charles Emmett Maher}
\affiliation{Department of Chemistry, Princeton University, Princeton, NJ 08544, United States of America}
\author{Yang Jiao}
\affiliation{Materials Science and Engineering, Arizona State University, Tempe, AZ 85287, United States of America}
\affiliation{Department of Physics, Arizona State University, Tempe, AZ 85287, United States of America}
\author{Salvatore Torquato}
\affiliation{Department of Chemistry, Princeton University, Princeton, NJ 08544, United States of America}
\affiliation{Department of Physics, Princeton University, Princeton, NJ 08544, United States of America}
\affiliation{Princeton Materials Institute, Princeton University, Princeton, NJ 08544, United States of America}
\affiliation{Program in Applied and Computational Mathematics, Princeton University, Princeton, NJ 08544, United States of America}


\begin{abstract}
The maximally random jammed (MRJ) state is the most random (i.e., disordered) configuration of strictly jammed (mechanically rigid) nonoverlapping objects.
MRJ packings are hyperuniform, meaning their long-wavelength density fluctuations are anomalously suppressed compared to typical disordered systems, i.e., their structure factors $S(\mathbf{k})$ tend to zero as the wave number $|\mathbf{k}|$ tends to zero.
Here, we show that generating high-quality strictly jammed states for Euclidean space dimensions $d = 3,4,$ and $5$ is of paramount importance in ensuring hyperuniformity and extracting precise values of the hyperuniformity exponent $\alpha > 0$ for MRJ states, defined by the power-law behavior of $S(\mathbf{k})\sim|\mathbf{k}|^{\alpha}$ in the limit $|\mathbf{k}|\rightarrow0$.
Moreover, we show that for fixed $d$ it is more difficult to ensure jamming as the particle number $N$ increases, which results in packings that are nonhyperuniform.
Free-volume theory arguments suggest that the \textit{ideal} MRJ state does not contain rattlers, which act as defects in numerically generated packings.
As $d$ increases, we find that the fraction of rattlers decreases substantially.
Our analysis of the largest truly jammed packings suggests that the \textit{ideal} MRJ packings for all dimensions $d\geq3$ are hyperuniform with $\alpha = d - 2$, implying the packings become more hyperuniform as $d$ increases. 
The differences in $\alpha$ between MRJ packings and recently proposed Manna-class random close packed (RCP) states, which were reported to have $\alpha = 0.25$ in $d=3$ and be nonhyperuniform ($\alpha = 0$) for $d = 4$ and $d = 5$, demonstrate the vivid distinctions between the large-scale structure of RCP and MRJ states in these dimensions.
Our paper clarifies the importance of the link between true jamming and hyperuniformity and motivates the development of an algorithm to produce rattler-free three-dimensional MRJ packings.
\end{abstract}

\maketitle

\section{Introduction}
Disordered hyperuniform many-particle systems in $d$-dimensional Euclidean space $\mathbb{R}^d$ are emerging exotic amorphous states of matter in which fluctuations are anomalously suppressed at infinite wavelengths compared to ordinary liquids \cite{03HU, HU_PhysRep}.
Such unusual disordered systems play a vital role in a variety of fields and contexts, including wave transport in complex media \cite{Fr17, HU_PhysRep, Oh22, Gr23, ref31, torquato2022extraordinary}, structural glasses and supercooled liquids \cite{Ma22, Ya23}, vortices in superconductors \cite{VortPlus, Sa23}, random matrix theory \cite{HU_PhysRep, Fo23}, quantum systems \cite{HU_PhysRep, Go22, Ab23}, and biological systems \cite{Hu21, Wi23}, among many other examples.
A \textit{hyperuniform} many-particle system at number density $\rho$ is one in which the local number variance $\sigma_N^2(R)\equiv\langle N(R)^2\rangle - \langle N(R)\rangle^2$ of particles within a spherical observation window of radius $R$, grows in the large-$R$ limit slower than $R^d$, where angular brackets denote an ensemble average \cite{03HU, HU_PhysRep}.
Equivalently, a many-particle system is hyperuniform if the structure factor $S(\mathbf{k})$ tends to zero as the wave number $k\equiv|\mathbf{k}|$ tends to zero, i.e.,
\begin{equation}
    \lim_{|\mathbf{k}|\rightarrow0}S(\mathbf{k})=0.
\end{equation}
Disordered packings in $\mathbb{R}^d$, which are collections of nonoverlapping particles, are important models in, e.g., soft matter \cite{SalBook,hughel_1965,zallen_2007,SMEX_2, GOcite, MRJ_defn}, materials science \cite{SalBook, mehta_1994, MDcite}, and biology \cite{BioEX1,BioEX2,BioEX3} (see, Refs. [\citenum{PackingRev}, \citenum{Torquato_PackingPersp}], and references therein for a more exhaustive list of examples).
Disordered packings of spheres have been shown to exhibit hyperuniformity when reaching certain putatively jammed (mechanically rigid) states \cite{Zachary_QLRlet,Zachary_QLRI, BRO_origin, BRO_highdim, Corwin_Jammed,Parisi_QLR,MRJ_CritSlow,Donev_1m,HU_USGlass,Weekscite}.

Torquato and Stillinger \cite{03HU} conjectured that infinitely large packings of identical frictionless hard spheres that are saturated and strictly jammed are hyperuniform.
A packing is strictly jammed if there is no possible collective rearrangement of some finite subset of particles, and no volume non-increasing deformation that can be applied to the packing without violating the impenetrability constraints of the particles. 
A packing is saturated if there is no available space to add another particle of the same type to the packing without resulting in interparticle overlaps. 
It has previously been reported that maximally random jammed (MRJ) packings of identical spheres \textcolor{black}{are hyperuniform} \cite{03HU, Donev_1m, MRJ_HopkinsGlass, MRJ_CritSlow, MRJ_Static, MRJ_HighDim, MRJ_Superballs}.
The MRJ states of several other particle geometries \cite{MRJ_Superballs, cinacchi2022dense, MRJ_platonic, Chen_MRJtruntet, Donev_MRJelip, Jiao_MRJballs} have also been characterized. 
The MRJ packing state is the most disordered packing (as measured by some set of scalar order metrics) that is also subject to strict jamming \cite{MRJ_defn}. 
The ideal MRJ state is rattler free, which is expected to yield perfect hyperuniformity and the most disordered strictly jammed state \cite{MRJ_CritSlow, Sal_AB}.
This notion is supported by free-volume theory arguments \cite{Sal_AB}, which predict that $S(0) = 0$ at the jammed state in the absence of defects (i.e., rattlers).

A stringent test of the conjecture for MRJ packings requires generation of large packings to access the long-range (small wave number) behavior, while maintaining strict jamming.
For this purpose, Donev {\it  et al.} \cite{Donev_1m} generated three-dimensional (3D) MRJ packings with $N = 10^6$ monodisperse spheres using a modified Lubachevsky-Stillinger (LS) event-driven algorithm \cite{Donev_DTSAlgo} and by fitting the data on a linear scale, concluded they were hyperuniform with $\alpha = 1$. 
Here, $\alpha$ is the power-law scaling of the small-wave number behavior of $S(\mathbf{k})$, specifically, $S(\mathbf{k})\sim |\mathbf{k}|^{\alpha}$ as $|\mathbf{k}|\rightarrow 0$. 
However, as we show below, these packings are not truly strictly jammed, which influences the value of $\alpha$ extracted.
In recent work, Wilken et al. \cite{BRO_origin} found that fitting $S(k)$ for the packing from Ref. [\citenum{Donev_1m}] on a logarithmic scale reveals a substantially lower $\alpha$ ($\sim0.25$).
Indeed, we also find that fitting this $S(k)$ on a logarithmic scale results in a smaller value of $\alpha$ compared to a direct fit of the linear data (cf. Fig. 6).
Moreover, previous studies have found that sphere packings with $N\gtrsim 1000$ produced using the LS algorithm are unable to attain isostaticity \cite{MRJ_PairStat,MRJ_Superballs}. 
In addition, work from Skoge {\it  et al.} \cite{MRJ_HighDim} suggests that four-dimensional (4D) and five-dimensional (5D) MRJ packings generated via an LS algorithm are also hyperuniform, but, similarly to Donev {\it  et al.} [\citenum{Donev_1m}], did not ensure strict jamming.
\textcolor{black}{It is noteworthy that very tiny collective particle displacements in a nonhyperuniform system can lead to hyperuniform systems \cite{HU_PhysRep}, dispelling any notion that similarities in local statistics always implies similarities in global (hyperuniform) statistics. This example lends additional support to our hypothesis that the small changes in the local packing structure that occur in the vicinity of the strict jamming point can drastically impact the hyperuniformity and precise value of $\alpha$ for a given packing.}   

Hyperuniform many-particle systems are poised at a critical point where large-scale density fluctuations are anomalously suppressed such that the {\it direct correlation function} is long-ranged \cite{03HU, HU_PhysRep}.
While packings can often be generated that have a high reduced pressure and hence appear jammed by conventional tests, rigorous jamming tests reveal that they often are not due to a type of ``critical slowing down'', and thus have deviations from perfect hyperuniformity \cite{MRJ_CritSlow}.
Critical slowing down in this context means the packing's rearrangements in configuration space become locally confined by high-dimensional ``bottlenecks'' from which escape is a rare event.
Such bottlenecks become increasingly more difficult to escape from as the size of the packing increases \cite{MRJ_CritSlow}.
    
Current numerically produced MRJ-like sphere packings invariably have a small concentration of \textit{rattlers} $\phi_R$ \cite{TJ_2D, Rattler_Char, Kansal_Diversity, Jiao_Diversity, MRJ_HighDim}, which are particles that are not jammed but are locally imprisoned by their neighbors, which decreases with dimension \cite{MRJ_HighDim}.
The remainder of the jammed spheres are referred to as the \textit{backbone}.
The value of $\phi_R$ for a given $d$ is also known to be affected by the protocol used for jamming \cite{TJ_Alg}.
While one would not generally expect exact hyperuniformity for disordered packings with rattlers, we have shown that when jamming is ensured, the packings are very nearly hyperuniform, and deviations from hyperuniformity correlate with an inability to ensure jamming, suggesting that strict jamming and hyperuniformity are indeed linked \cite{MRJ_CritSlow}.
       
Herein, we endeavor to more carefully study and characterize the hyperuniformity of MRJ sphere packings by generating and analyzing large MRJ sphere packings in $d=3,4,5$ with more stringent jamming criteria than Refs. [\citenum{Donev_1m, MRJ_HighDim}].
To generate these packings, we employ a hybrid scheme where we first produce a nearly jammed sphere packing using the LS algorithm, which is used as an initial condition for the Torquato-Jiao (TJ) linear programming (LP) jamming algorithm \cite{TJ_Alg}, which is known to reliably produce strictly jammed packings \cite{TJ_binary, TJ_2D,Rattler_Char, MRJ_Static}.

To characterize the extent to which the hybrid scheme can bring a packing of $N$ identical hyperspheres to strict jamming in $\mathbb{R}^3,\mathbb{R}^4,\mathbb{R}^5$ and how this relates to ensuring hyperuniformity, we execute the following steps:
\begin{enumerate}
    \item We generate ensembles of 3D, 4D, and 5D hypersphere packings with different $N$ and show that as $N$ increases, ensuring jamming becomes more difficult, which results in a degradation of the hyperuniformity of the ensemble.
    \item For a fixed $N$ and $d$, we show that as the jamming packing fraction $\phi_J$ is approached, both the quality of the jammed backbone and degree of hyperuniformity improve.
    \item In each spatial dimension $d = 3,4,5$, we use these results to determine the largest $N$ that reliably yields strictly jammed packings and characterize their hyperuniformity.
    \item We use rigorous LP jamming tests \cite{MRJ_CritSlow, LPTest_Donev} on 3D MRJ packings with $N = 2500$ to quantify the degree to which these packings are strictly jammed and demonstrate that the hybrid scheme reliably produces strictly jammed packings.
\end{enumerate}

Subsequently, we compute $S(k)$ of these large high-quality jammed packings to extract $\alpha$.
We first directly fit the small-$k$ behavior of $S(k)$ on a logarithmic scale to determine the power-law scaling in the vicinity of the origin.
The diffusion spreadability, which is a dynamical probe that directly links time-dependant diffusive transport with the microstructure of a two-phase medium \cite{Spread_Origin}, has been used profitably to extract the exponent $\alpha$ from a variety of hyperuniform and nonhyperuniform systems \cite{MRJ_Superballs, Spread_Algo, Spread_Wang, Spread_Origin}.
Here, we fit the long-time behavior of the \textit{excess spreadability} $\mathcal{S}(\infty)-\mathcal{S}(t)$ to a power law of the form $\mathcal{S}(\infty)-\mathcal{S}(t) \sim t^{(d+\alpha)/2}$ to extract $\alpha$ from these packings.

We find that the quality of the jammed backbone of 3D MRJ sphere packings increases sharply in the vicinity of the strict jamming point and the $S(k)$ has a concomitant decrease in the magnitude of its small-$k$ behavior, indicating an increase in hyperuniformity.
The largest 3D packing size that reliably reaches strict jamming is $N = 5000$, and packings with larger $N$ have decreased backbone quality and degraded hyperuniformity.
We find that the $N = 5000$ 3D MRJ packings have a scaling exponent $\alpha = 0.973\pm0.02$ and $\phi_R=1.6\%$.
This $\alpha$ is slightly smaller than the expected $\alpha = 1$ \cite{HU_PhysRep}, which we surmise is due to the nonzero concentration of rattlers.
Moreover, this value of $\alpha$ is larger than those extracted from the logarithmic fitting analysis of the $S(k)$ given by Donev {\it et al.} \cite{Donev_1m} reported in Ref. [\citenum{BRO_origin}] and Fig 6.
This clearly demonstrates the importance of achieving strict jamming to the hyperuniformity scaling of a packing.

For 4D and 5D MRJ packings, we find that the backbone quality and degree of hyperuniformity are more susceptible to changes in the degree of strict jamming when compared to the 3D packings.
This suggests that achieving strict jamming is more important to ensuring hyperuniformity as $d$ increases.
For the largest 4D ($N = 10000$) and 5D ($N = 20000$) strictly jammed packings, we find $\alpha = 1.946\pm0.013$ and $2.926\pm0.12$, respectively.
As will be explained in Sec. V B, these values suggest the scaling relation $\alpha = d - 2$ for $d\geq3$.
Each of these $\alpha$ values are greater than those recently reported for 3D \cite{BRO_origin}, 4D, and 5D \cite{BRO_highdim} random close packed (RCP) states, which vividly demonstrates that the jamming requirement differentiates the MRJ state from RCP states.

\section{Background}
\subsection{Pair statistics}
Systems of point particles in $\mathbb{R}^d$ are fully spatially characterized by an infinite set of $n$-particle correlation functions $\rho_n(\mathbf{r}_1,\dots,\mathbf{r}_n)$, which are proportional to the probability of finding $n$ particles at the positions $\mathbf{r}_1,\dots,\mathbf{r}_n$ \cite{DecorPrinc}.
For statistically homogeneous systems, $\rho_1(\mathbf{r}_1)=\rho$, and $\rho_2(\mathbf{r}_1,\mathbf{r}_2)=\rho^2 g_2(\mathbf{r})$, where $\mathbf{r}=\mathbf{r}_1-\mathbf{r}_2$, and $g_2(\mathbf{r})$ is the pair correlation function.
If the system is also statistically isotropic, then $g_2(\mathbf{r})=g_2(r)$, where $r=|r|$.
The ensemble-averaged structure factor $S(\mathbf{k})$ is defined as
\begin{equation}
    S(\mathbf{k})=1+\rho\Tilde{h}(\mathbf{k})
\end{equation}
where $\Tilde{h}(\mathbf{k})$ is the Fourier transform of the total correlation function $h(\mathbf{r})=g_2(\mathbf{r})-1$.

For a single periodic point configuration with $N$ particles at positions $\mathbf{r}^N = (\mathbf{r}_1,\dots,\mathbf{r}_N)$ within a fundamental cell $F$ of a lattice $\Lambda$, the scattering intensity $\mathbb{S}(\mathbf{k})$ is given by

\begin{equation}\label{eq:Skcomp}
    \mathbb{S}(\mathbf{k}) = \frac{|\sum_{j=1}^N \textrm{exp}(-i\mathbf{k}\cdot\mathbf{r}_j)|^2}{N}.
\end{equation}
In the thermodynamic limit, an ensemble of $N$-particle configurations in $F$ is related to $S(\mathbf{k})$ by
\begin{equation}
    \lim_{N,V_F\rightarrow\infty}\langle \mathbb{S}(\mathbf{k})\rangle = (2\pi)^d \rho \delta(\mathbf{k}) + S(\mathbf{k}),
\end{equation}
where $V_F$ is the volume of the fundamental cell and $\delta$ is the Dirac delta function \cite{03HU}.
For finite-$N$ simulations under periodic boundary conditions, Eq. (\ref{eq:Skcomp}) is used to compute
 $S(\mathbf{k})$ directly by averaging over configurations.

One can treat packings as two-phase heterogeneous media by considering the space between the particles as the matrix phase $\mathcal{V}_1$ and the particles themselves as phase two $\mathcal{V}_2$ \cite{PackingTwoPhase}.
The packing microstructure can be fully characterized by a countably infinite set of $n$-point probability functions $S_n^{(i)}(\mathbf{x}_1,\dots,\mathbf{x}_n)$, defined by \cite{SalBook}
\begin{equation}
    S_n^{(i)}(\mathbf{x}_1,\dots,\mathbf{x}_n) = \left\langle \prod_{j=1}^n\mathcal{I}^{(i)}(\mathbf{x}_j)\right\rangle, 
\end{equation}
where $\mathcal{I}^{(i)}$ is the indicator function for phase $i$:
\begin{equation}
    \mathcal{I}^{(i)}(\mathbf{x}) =
    \begin{cases}
    1, & \mathbf{x} \in \mathcal{V}_i\\
    0, & \textrm{else}.
    \end{cases}
\end{equation}
The functions $S_n^{(i)}(\mathbf{x}_1,\dots,\mathbf{x}_n)$ give the probability of finding $n$ points at positions $(\mathbf{x}_1,\dots,\mathbf{x}_n)$ in phase $i$.
Hereafter, we drop the superscript $i$ and restrict our discussion to the particle phase $\mathcal{V}_2$.

For statistically homogeneous media, $S_n(\mathbf{x}_1,\dots,\mathbf{x}_n)$ is translationally invariant and, in particular, $S_1$ is independent of position and equal to the packing fraction $\phi$,
while the two-point correlation function $S_2(\mathbf{r})$ depends on the displacement vector $\mathbf{r}\equiv\mathbf{x}_2-\mathbf{x}_1$.
The corresponding two-point autocovariance function $\chi_{_V}(\mathbf{r})$ \cite{SalBook,StochaticGeo_Text, Quint_Autoco} is obtained by subtracting the long-range behavior from $S_2(\mathbf{r})$:
\begin{equation}
    \chi_{_V}(\mathbf{r})=S_2(\mathbf{r})-\phi^2.
\end{equation}
The nonnegative \textit{spectral density}, $\spD{\mathbf{k}}$, is defined as the Fourier transform of $\chi_{_V}(\mathbf{r})$ \cite{SalBook}.
For a monodisperse packing of spheres with radius $a$, it is known that \cite{Torquato_SD1, SalBook, Torquato_DisorderHUHet}
\begin{equation}
    \spD{\mathbf{k}}=\rho|\tilde{m}(\mathbf{k};a)|^2S(\mathbf{k}),
\end{equation}
where $\tilde{m}(\mathbf{k};a)$ is the Fourier transform of the particle indicator function (form factor) defined as
\begin{equation}
    m(\mathbf{r};a) =
    \begin{cases}
    1, & |\mathbf{r}| < a\\
    0, & \textrm{otherwise,}
    \end{cases}
\end{equation}
where $\mathbf{r}$ is a vector measured with respect to the particle centroid.

For single finite configurations of $N$ identical hard spheres under periodic boundary conditions in $F$ with volume $V_F$, $\spD{\mathbf{k}}$ can be expressed as \cite{Zachary_QLRI}
\begin{equation}
\begin{split}\label{eq:spD}
    \spD{\mathbf{k}} = &\frac{\left|\sum_{j=1}^N\textrm{exp}(-i\mathbf{k}\cdot\mathbf{r}_j)\tilde{m}(\mathbf{k};a)\right|^2}{V_F}\\
    &(\mathbf{k}\neq0),    
\end{split}
\end{equation}
where $\{\mathbf{r}_\mathit{j}\}$ denotes the set of particle centroids. 
Equation (\ref{eq:spD}) is used to compute the spectral densities of the MRJ sphere packings.

\subsection{Strict-jamming criteria}
A \textit{jammed} hard particle packing is one in which each particle makes contact with its neighbors in such a way that certain levels of mechanical stability are conferred to the packing \cite{PTest_1}.
Such packings can be placed into three mathematically precise categories based on the type of mechanical stability conferred, which in order from least to most stable are \cite{PackingRev, PTest_1}:
(1) \textit{Local jamming}: no individual particle can be moved while holding all other particles fixed.
(2) \textit{Collective jamming}: the packing is locally jammed, and no collective motion of a finite subset of particles is possible.
(3) \textit{Strict jamming}: the packing is collectively jammed and all volume-nonincreasing deformations are disallowed by the impenetrability constraint.

Of particular interest are MRJ packings of spheres under the strict jamming constraint.
Such disordered states can be viewed as prototypical glasses because they are maximally disordered, perfectly rigid (have infinite elastic moduli), and perfectly nonergodic \cite{PackingRev, MRJ_defn}.
MRJ sphere packings are known to be isostatic \cite{MRJ_PairStat, Isostat_Ohern}, which means that the total number of interparticle contacts (constraints) in the packing is equal to the total number of degrees of freedom and that all of the constraints are linearly independent.
For a finite, rattler-free, sphere packing in a periodic deformable simulation box to be strictly jammed, the total number of interparticle contacts $Z(N)$ must be \cite{MRJ_PairStat,TJ_binary}
\begin{equation}\label{eq:iso_strict}
    Z(N) = d(N-1)+d(d+1)/2.
\end{equation}
Equivalently, each sphere must have $2d-2d/N+d(d+1)/N$ contacts on average, which is the minimal number of contacts to constrain all degrees of freedom available to the particles.

In practice, disordered jammed packings contain a small concentration of \textit{rattlers}.
Monodisperse 3D sphere packings generated using the TJ algorithm have $\phi_R \sim1.5\%$ \cite{Rattler_Char} and early results indicate the rattler fraction decreases as $d$ increases \cite{TJ_Alg}.
Jamming precludes the existence of rattlers \cite{PackingRev, PTest_1}.
Nevertheless, it is the significant majority of particles that confers mechanical rigidity to the packing, and in any case, the rattlers could be removed (in computer simulations) without disrupting the remaining jammed backbone \cite{PackingRev,Torquato_PackingPersp}.

To rigorously test if a frictionless sphere packing is strictly jammed, Donev {\it et al.} \cite{LPTest_Donev} introduced a method that uses randomized sequential LP.
The algorithm maximizes the work done by random body forces on spheres in the packing displacing them and straining the simulation box while obeying the constraints that the spheres do not overlap and the volume of the box does not increase.
Atkinson {\it et al.} \cite{MRJ_CritSlow} improved on this scheme by instead choosing body forces that are more likely to induce local rearrangements of particles, and found that the LP scheme can more accurately determine if a packing is jammed than standard methods (e.g., the pressure leak test \cite{PTest_1,LPTest_Donev,PTest_2,PTest_3}).

For packings with ideal contacts, i.e., ones where interparticle contacts are exact, any sphere that moves is considered a rattler and if every sphere is a rattler, then the packing is unjammed.
Numerically produced packings, however, cannot have true contacts and thus one has to account for the fact that even the spheres comprising the jammed backbone will be able to move a small amount.
Therefore, we instead consider the degree to which a packing is strictly jammed by \textcolor{black}{examining the maximum Euclidean distance a backbone sphere moves from its original location when} these body forces are applied \textcolor{black}{over the course of the algorithm described in Ref. [\citenum{MRJ_CritSlow}]}.

Previous studies have also judged how well-jammed a packing is by examining the plateau in the cumulative coordination function $Z(r)$, a quantity related to the integral of the pair correlation function $g_2(r)$ \cite{MRJ_PairStat,BRO_origin}, or similarly, the gap between the contact tolerances which result in an exactly isostatic jammed backbone and a backbone with a single additional contact \cite{Rattler_Char}. 
While these measures correctly determine the smallest contact tolerance associated with an isostatic, strictly jammed backbone they often underestimate the size of the largest gap because detection of spurious interparticle contacts between the backbone spheres make the packing appear hyperstatic. 
Such additional intrabackbone contacts are found at smaller contact tolerances as the packing size increases \cite{Rattler_Char}, and thus $Z(r)$ is not useful when assessing the quality of large jammed sphere packings. 

\subsection{Hyperuniformity}
Consider systems characterized by a structure factor with a radial power law in the vicinity of the origin: 
\begin{equation}
    S(\mathbf{k})\sim|\mathbf{k}|^{\alpha}\;\textrm{for}\;|\mathbf{k}|\rightarrow0.
\end{equation}
For hyperuniform systems, $\alpha > 0$, and its specific value determines three different large-$R$ scaling behaviors or classes of the number variance [\cite{03HU,HU_PhysRep,Zachary_HU}]
\begin{equation}\label{eq:classes}
    \sigma^2_N(R)\sim
    \begin{cases}
    R^{d-1}&\alpha > 1, \textrm{class I}\\
    R^{d-1}\textrm{ln}(R)&\alpha = 1, \textrm{class II}\\
    R^{d-\alpha}&\alpha < 1, \textrm{class III}.
    \end{cases}
\end{equation}
Classes I and III describe the strongest and weakest forms of hyperuniformity, respectively.
Such classes apply analogously to spectral densities possessing a radial power-law form in the vicinity of the origin \cite{Torquato_DisorderHUHet}, i.e.,
\begin{equation}
    \spD{\mathbf{k}}\sim|\mathbf{k}|^{\alpha}\;\textrm{for}\;|\mathbf{k}|\rightarrow0.
\end{equation}
These classes instead correspond to a large-$R$ scaling behavior \textit{volume-fraction} variance, which considers the volume fraction of a desired phase in an observation window of radius $R$.
\textcolor{black}{The precise scaling form (14) contributes to the determination of diffusion spreadability \cite{Spread_Origin}, nuclear magnetic resonance and magnetic resonance imaging measurements \cite{mag1,mag2,mag3}, rigorous upper bounds on the fluid permeability of hyperuniform two-phase media \cite{AWR}, the electromagnetic wave characteristics of composites beyond the quasistatic regime \cite{qs1}, photonic bandgaps in disordered media \cite{pbg1}, and the optical transparency of 1D disordered media \cite{1Dopt}.}

To properly ascertain the hyperuniformity of a packing, one must run simulations with a large number of particles $N$ to access the small-$k$ regime of $S(k)$ or $\spD{k}$.
This is complicated by the nosiness of small-$k$ data, numerical and protocol-dependent errors, and reliance on extrapolations of such uncertain data to the $k\rightarrow0$ limit \cite{MRJ_CritSlow}.
It is also known, however, that the ``critical slowing down'' on approach to jammed states becomes more difficult to overcome as $N$ increases because the complex collective motions required for jamming become practically impossible \cite{MRJ_CritSlow, Crit_Book1, Crit_Book2}.
Thus, we generate putatively MRJ packings with a range of $N$ to determine the largest size at which we can still ensure high-quality strict jamming.

\subsection{Spreadability}
Recent work \cite{Spread_Origin,Spread_Algo,Spread_Wang,MRJ_Superballs} has revealed that the time-dependent \textit{spreadability} is a powerful dynamic-based figure of merit to probe and classify the spectrum of possible microstructures of two-phase media across length scales.
Consider the time-dependent problem describing the mass transfer of a solute between two phases and assume that the solute is initially only present in one phase, specifically the particle phase, and both phases have the same $\mathcal{D}$.
The fraction of total solute present in the void space as a function of time $\mathcal{S}(t)$, is termed the \textit{spreadability} because it is a measure of the spreadability of diffusion information as a function of time.
Qualitatively, given two different two-phase systems at some time $t$, the one with a larger value of $\mathcal{S}(t)$ spreads diffusion information more rapidly.
Recently, Torquato showed that the \textit{excess spreadability} $\mathcal{S}(\infty)-\mathcal{S}(t)$ can be expressed in Fourier space in any dimension $d$ as \cite{Spread_Origin}:
\begin{equation}\label{eq:FT_spread}
    \mathcal{S}(\infty)-\mathcal{S}(t) = \frac{d\omega_d}{(2\pi)^d \phi}\int^{\infty}_0 k^{d-1}\spD{k}\textrm{exp}[-k^2\mathcal{D}t]dk,
\end{equation}
where $\omega_d$ is the volume of a $d$-dimensional unit sphere:
\begin{equation}
    \omega_d=\frac{\pi^{d/2}}{\Gamma(1+d/2)}.
\end{equation}

In the particular case of two-phase media with $\spD{k}$ that obeys a power-law scaling in the $k\rightarrow0$ limit,
\begin{equation}
    \lim_{k\rightarrow 0} \spD{k} = B|k\ell|^{\alpha},
\end{equation}
where $B$ is a positive dimensionless constant, $\ell$ represents some characteristic microscopic length scale, and $\alpha\in(-d,\infty)$, the long-time behavior of the excess spreadability can be written as \cite{Spread_Origin}
\begin{equation}\label{eq:PLF}
    \mathcal{S}(\infty)-\mathcal{S}(t) \sim 1/t^{(d+\alpha)/2}.
\end{equation}
Thus, one can use the spreadability to ascertain and classify the hyperuniformity of such two-phase media.
Here, we use Eq. (\ref{eq:FT_spread}) to compute $\mathcal{S}(\infty)-\mathcal{S}(t)$ for the sphere packings and fit the long-time behavior using the algorithm detailed in Ref. [\citenum{Spread_Wang}].
\textcolor{black}{We note that, unlike direct fitting methods, the spreadability does not require a choice of wave numbers to fit over, nor a fit of a chosen form, and thus is less susceptible to errors introduced by these choices \cite{Spread_Origin, Spread_Wang, MRJ_Superballs, Spread_Algo}.}

\section{Packing Generation}
To produce the putatively MRJ packings examined here, we start by randomly placing nonoverlapping (hyper)spheres in a (hyper)cubic simulation box with an initial packing fraction $\phi_i = 0.2\; (d = 3), 0.1\; (d = 4), 0.05\; (d = 5)$.
We then use the implementation of the LS algorithm given in Ref. [\citenum{Donev_DTSAlgo}] to bring these packings to a reduced pressure of 75 at a dimensionless expansion rate of $\Gamma = d\mathbb{D}(t)/dt\sqrt{m/(k_BT)}= 0.1$, where $m$ is the mass of the \textcolor{black}{sphere, and $\mathbb{D}(t)$ is the time-dependent sphere diameter. The particle diameter at the end of the LS simulation is denoted by $D$.}
This densified, but unjammed, state is then used as an initial condition in the TJ algorithm.
A detailed description of the sequential LP algorithm, which solves the adaptive shrinking-cell problem, can be found in Ref. [\citenum{TJ_Alg}].
Here, we use an influence sphere size $\gamma = D/40$, a maximum translation magnitude of $\Delta x\leq D/200$, and a maximum box strain magnitude of $\epsilon\leq D/200$.
Simulations are terminated when the packing fraction $\phi$ decreases less than $10^{-15}$ over the course of two LP steps.
\textcolor{black}{Note that the specific choice of the terminal pressure chosen for the LS algorithm above does not significantly impact the fraction of rattlers after the execution of the TJ algorithm.}
We generated ensembles of 50 $N=2500$ packings, 50 $N = 5000$ packings, 50 $N = 16000$, and 10 $N = 25000$ packings in $d=3$; and generated ensembles of configurations for $N=5000$, $N=10000$, $N=20000$, and $N=25000$ packings in $d=4$, and ensembles of configurations for $N=10000$, $N=20000$, and $N=50000$ packings in $d=5$.
At the termination criterion, however, 4 $N = 16000$ 3D packings did not reach jamming and were left out of ensemble averaged results.
Additionally, we note here that Artiaco {\it et al.}, introduced another LP jamming algorithm in 2022 called \textit{Chain of Approximate Linear Programming for Packing Spherical Objects} (CALiPPSO) \cite{CALIPPSOpapers}. 
The key difference between TJ and CALiPPSO is that CALiPPSO forgoes shear deformations to improve efficiency and thus cannot produce the strictly jammed packings that we are interested in here.

\begin{figure*}
    \centering
     \subfloat[]{\hspace{-0.05\textwidth}\includegraphics[height=0.35\textwidth]{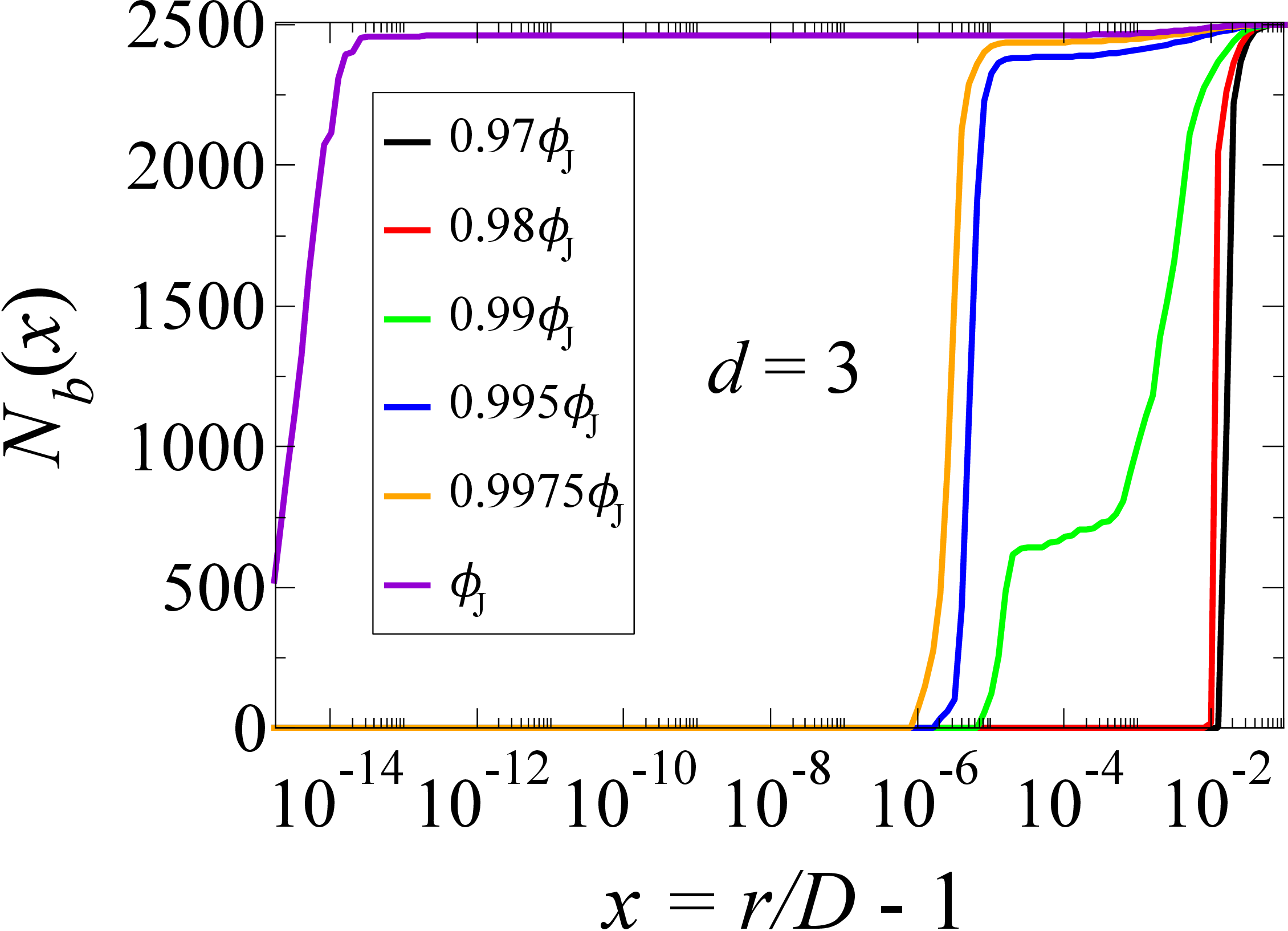}\hspace{0.05\textwidth}}
     \subfloat[]{\includegraphics[height=0.35\textwidth]{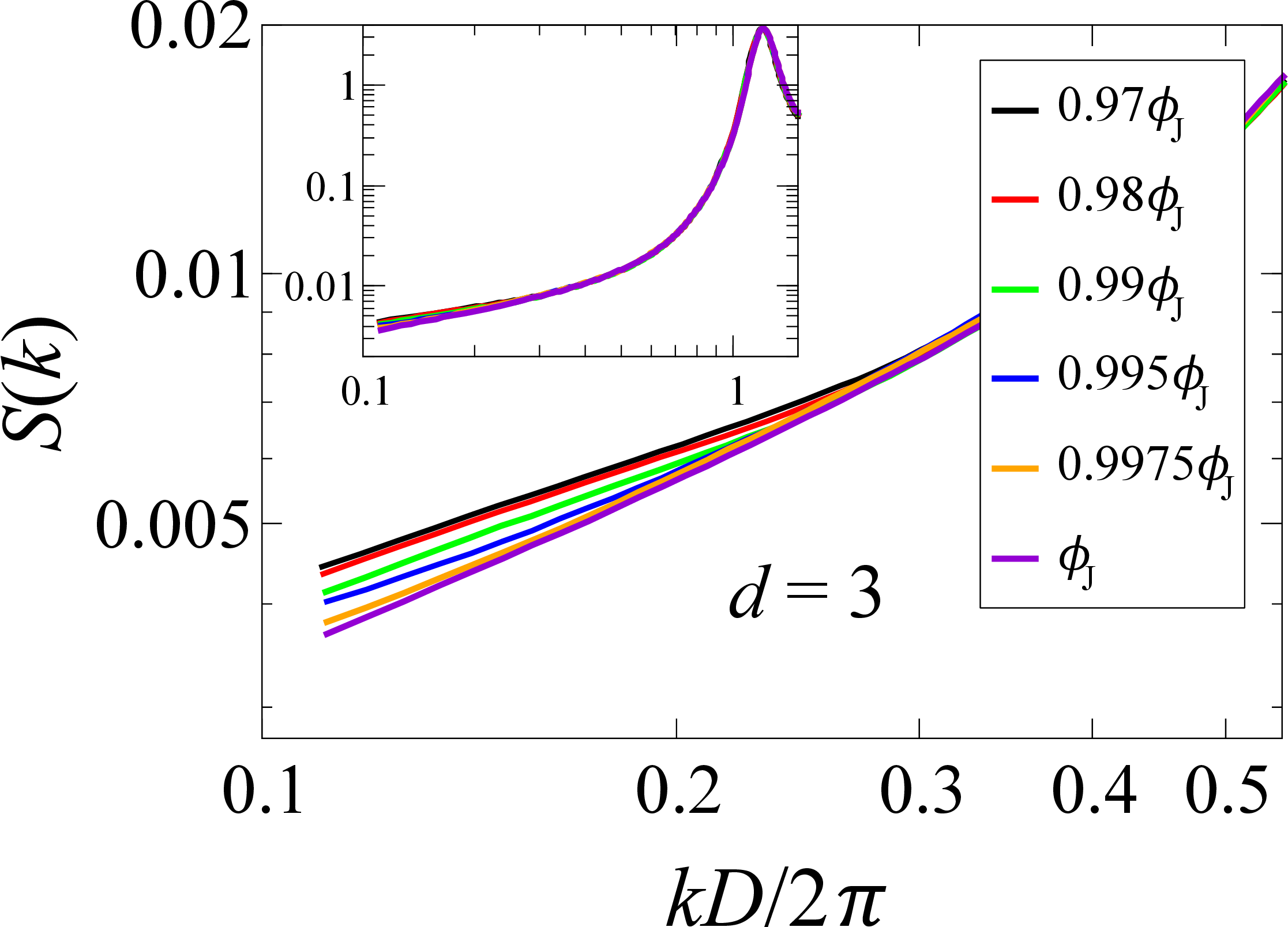}}
     \caption{The (a) number of backbone spheres $N_b(x)$ as a function of the scaled interparticle gap $r/D-1$, where $D$ is the particle diameter for several fractions of the jamming packing fraction $\phi_J$ = 0.638 and (b) corresponding structure factors $S(k)$ as a function of the scaled wave number $kD/2\pi$ in $\mathbb{R}^3$ with $N = 2500$. The inset of (b) shows the change in the structure factor occurs mostly on the large length scales, while the small- and intermediate-scale behavior does not change significantly.}
\end{figure*}

\section{Generation and Characterization of 3D MRJ Sphere Packings}
In the following subsections, we examine the relationship between hyperuniformity and proximity to jamming to show the importance of achieving strict jamming to ensure hyperuniformity and the precise determination of the exponent $\alpha$.
Additionally, we demonstrate the relationship between system size and jamming to find the largest packings for which we can ensure strict jamming.
To do this, we generate 3D MRJ spheres packings and conduct two different numerical experiments.
First, for a fixed $N$ we track the quality of the jammed backbone and the small-$k$ behavior of $S(k)$ as a function of $\phi$ as the strict jamming point is approached.
Then, we examine how these same two structural descriptors change as a function of the number of spheres $N$ for the putatively jammed states.
We also confirm that the TJ algorithm can reliably generate packings with putatively strictly jammed backbones using the LP-based pop jamming test \cite{MRJ_CritSlow}.

\subsection{Relationship between jamming quality and hyperuniformity}

\textcolor{black}{The TJ LP algorithm is formulated such that it will yield a strictly jammed packing to within a small numerical tolerance (see Ref. [\citenum{TJ_Alg}] for details about the formulation and tolerances of the algorithm).}
\textcolor{black}{We verify this claim by applying a rigorous LP jamming test to a set of putatively strictly jammed packings generated using the TJ algorithm.}
The LP jamming test introduced by Donev {\it et al.} \cite{LPTest_Donev}, while rigorous, requires a great deal of computational resources to carry out due to the random choice of applied body forces.
The pop test implementation of this algorithm by Atkinson {\it et al.} \cite{MRJ_CritSlow} increases the efficiency of this process by using a choice of body forces that are likely to result in a rearrangement \textcolor{black}{using a heuristic described in detail in Ref. [\citenum{MRJ_CritSlow}]}, which enables one to determine strict jamming with a high degree of accuracy.
We applied the pop test to an ensemble of 98 $N = 2500$ putatively MRJ sphere packings (with rattlers removed) generated using the hybrid scheme and tracked the average backbone translation from each applied body force.
We found that, at worst, a packing allowed an average backbone translation on the order of $10^{-6}D$, while the median value was on the order of $10^{-9}D$.
Thus, we have numerical evidence that the hybrid LS-to-TJ packing scheme reliably produces packings with high-quality strictly jammed backbones.

\begin{figure*}[!t]
    \centering
     \subfloat[]{\includegraphics[height=0.35\textwidth]{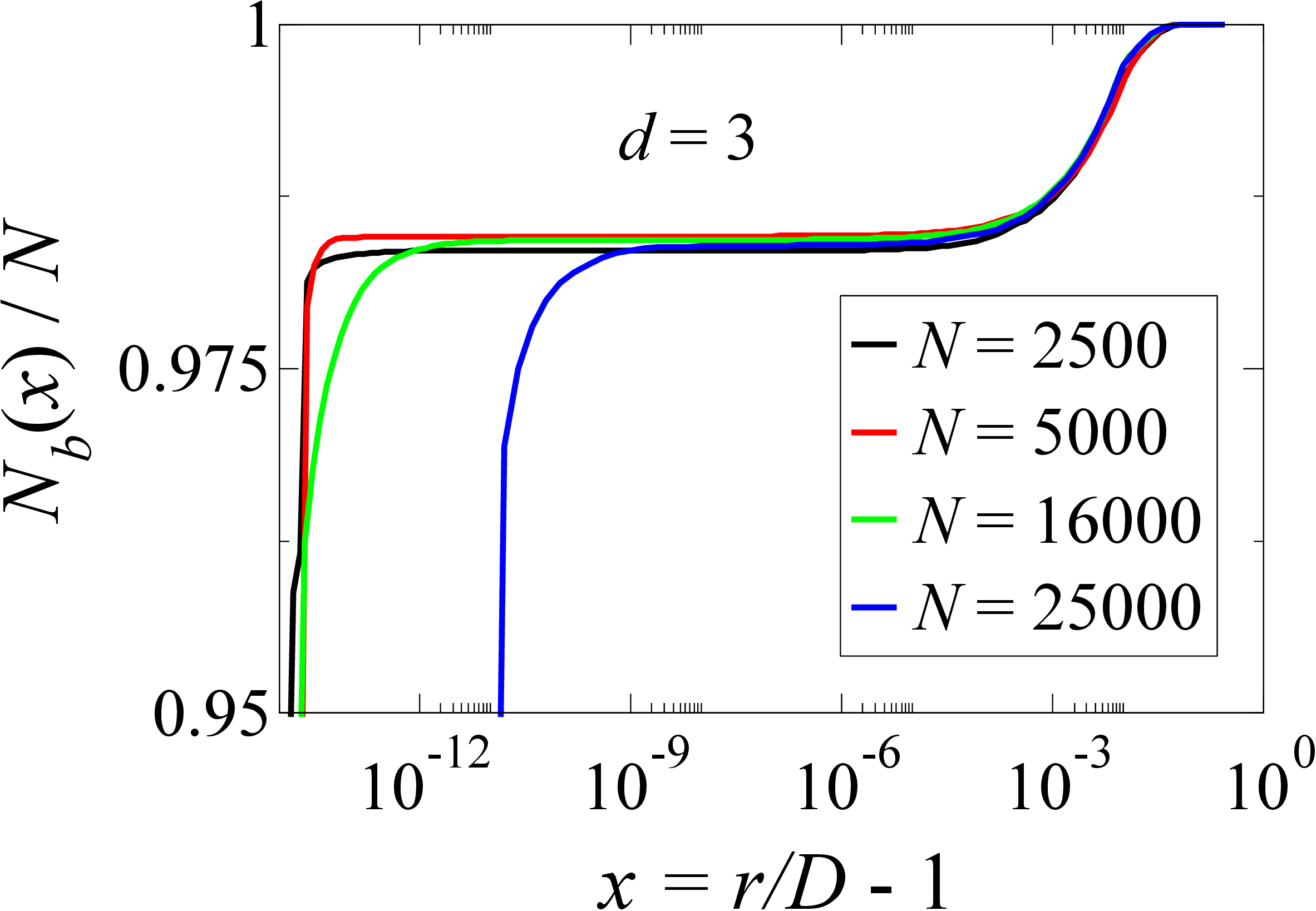}}
     \subfloat[]{\includegraphics[height=0.35\textwidth]{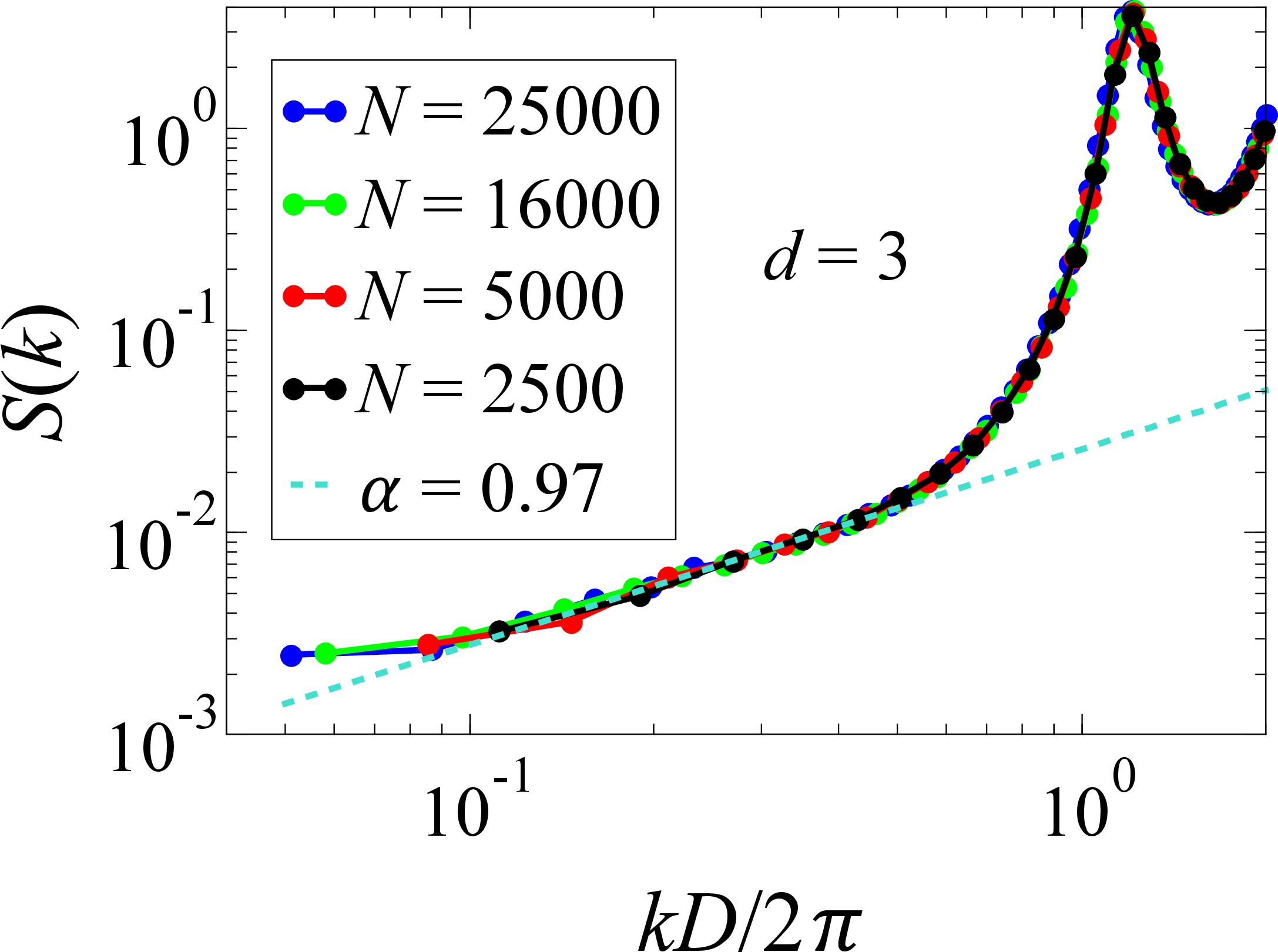}}
     \caption{The (a) number of backbone spheres $N_b(x)$ scaled by the number of spheres as a function of the scaled interparticle gap $r/D-1$, where $D$ is the particle diameter and the (b) corresponding structure factors $S(k)$ as a function of the scaled wave number $kD/2\pi$ in $\mathbb{R}^3$ for packings with $N = 2500$ ($\phi_J = 0.638$), $N = 5000$ ($\phi_J = 0.639$), $N = 16000$ ($\phi_J = 0.639$), and $N = 25000$ ($\phi_J = 0.639$). (b) has a dashed cyan line indicating $\alpha = 0.97$ scaling.}
\end{figure*}

To circumvent the issues stated earlier regarding the use of $Z(r)$ as a measure of jamming quality, we instead examine the number of backbone spheres $N_b(x)$ as a function of the contact tolerance $x = (r/D) - 1$. 
To compute $N_b(x)$, we determine all of the intersphere contacts at a given contact tolerance, iteratively remove the spheres which are rattlers, and count the remaining jammed spheres.
\textcolor{black}{In the calculation of $N_b(x)$, we use the local jamming definition for a rattler, i.e.: if a sphere has at least $d+1$ contacts that are not all on the same hemisphere of the particle, it is not a rattler \cite{dplus1}.}
\textcolor{black}{While this method will underestimate the total number of rattlers, a backbone sphere will never be mistaken for a rattler until the contact tolerance $x$ becomes smaller than the numerical tolerance of the TJ algorithm because this local definition is necessary, but not sufficient, to determine rattlers that are not collectively or strictly jammed.}
This measure is robust to the spurious additional contacts counted between the backbone spheres when $N$ becomes large and the large-gap tolerance behavior instead only changes when a rattler is erroneously considered a backbone sphere.
Figure 1(a) demonstrates that as the putative jamming packing fraction $\phi_J$ is approached, the plateau in $N_b(x)$ increases in width dramatically.
The $N_b(x)$ curve at jamming in Fig 1(a) (purple line) indicates higher quality jamming than previous studies of LS-derived packings, which are not isostatic \cite{MRJ_PairStat}, and \textcolor{black}{biased random organization (BRO)}-derived packings, which only have isostatic backbones over a small range of contact tolerances near $\sim 10^{-3}D$ \cite{BRO_origin, BRO_highdim}.
In these cases, the $Z(x)$ curves indicate the packings are far from the jamming point and thus hyperuniformity cannot be enforced \cite{Sal_AB}.
Moreover, this implies that the high-quality strict jamming shown via the pop test and the broad plateau in $N_b(x)$ are related.

This increase in backbone quality is accompanied by a monotonic decrease in the magnitude of the small-$k$ behavior of $S(k)$ as $\phi_J$ is approached [see Fig. 1(b)], i.e., the packings transition from nonhyperuniform to hyperuniform as jamming is approached. 
In addition, only the small-$k$ behavior of $S(k)$ is altered this close to jamming, which indicates that even small deviations from jamming affect the large- and intermediate-scale structure of packings.
Thus, we clearly establish that achieving strict jamming ensures hyperuniformity, which is very important for the precise determination of $\alpha$ in subsequent sections.
This increase in hyperuniform character on approach to jamming is corroborated by packings produced using only LS \cite{MRJ_HopkinsGlass, MRJ_Static} and only TJ \cite{MRJ_Static}.
\textcolor{black}{These findings imply that, even though large 3D sphere packings produced using only the LS algorithm \cite{Donev_1m} and 3D biased-random organization sphere packings \cite{BRO_origin} are similar in disorder and $\phi$ to the hybrid packings examined here, they can still have a value of $\alpha$ much less than 1 if they are not truly jammed.}
\textcolor{black}{As noted in the Introduction, previous work demonstrated that very tiny collective particle displacements in a nonhyperuniform system can lead to hyperuniform systems \cite{HU_PhysRep}, which supports the notion that the small changes in packing structure that occur in the vicinity of the strict jamming point can drastically impact the hyperuniformity and precise value of $\alpha$ for a given packing.}

\subsection{Relationship between system size and hyperuniformity}

The size of the largest strictly jammed packing we can produce is constrained by the performance of the TJ algorithm.
To determine the largest packing which can reliably achieve strict jamming and, thus, ensure hyperuniformity, we produced ensembles of $N = 2500, 5000, 16000,$ and $25000$ spheres using the same set of parameters that ensure high-quality jamming, while also being computationally feasible, and examined the behavior of their $N_b(x)$ [Fig. 2(a)] and $S(k)$ [Fig. 2(b)]. 
The different plateau heights occur due to small changes in the rattler fraction between the different ensembles of configurations.
We find that $N = 2500$ and 5000 have similar backbone qualities and, correspondingly, similar $S(k)$. 
In addition, the precise values of $\alpha$ we extract from the excess spreadability and direct fitting methods agree within error: the excess spreadability (direct fit) gives $\alpha  = 0.967\pm0.020$ (0.953) for the $N = 2500$ ensemble and $\alpha = 0.973\pm0.020$ (0.987) for the $N = 5000$ ensemble.
\textcolor{black}{For these direct fits, we choose a range of $k$ values such that removing or adding another $k$ value to either end of the fit range does not significantly change $\alpha$. Note that the precise value of $\alpha$ is sensitive to the exact choice of fitting range, and thus, direct fitting of $S(k)$ or $\Tilde{\chi}_{_V}(k)$ is a less robust way of extracting $\alpha$ compared to the spreadability.}
\textcolor{black}{The uncertainty in these $\alpha$ measurements is the standard deviation of the $\alpha$ values extracted from the excess spreadabilities for each packing individually.}
Note that such values of $\alpha$ are not only appreciably higher than the LS-or BRO- derived exponents (which have a lower strict-jamming quality) but neatly equal to unity.
We suggest that the deviation from unity is due to the presence of a small concentration of rattlers (i.e., defects) and so we expect that $\alpha = 1$ for ideal 3D MRJ states, which is consistent with the results reported in Ref. [\citenum{MRJ_Static}].
The backbone quality for the $N = 16000$ and $N = 25000$ packings is worse, which is evident from the shorter plateau in $N_b(x)$ and structure factors that indicate the packings are nonhyperuniform.
Thus, $N = 5000$ is the largest $N$ for which we have an ensemble of packings with high-quality strictly jammed backbones.
The results in Fig. 2 further demonstrate that hyperuniformity cannot be ensured for packings away from strict jamming, which is more problematic as $N$ increases.

 \begin{figure*}[!t]
        \centering
        \subfloat[]{\hspace{-0.05\textwidth}\includegraphics[height=0.35\textwidth]{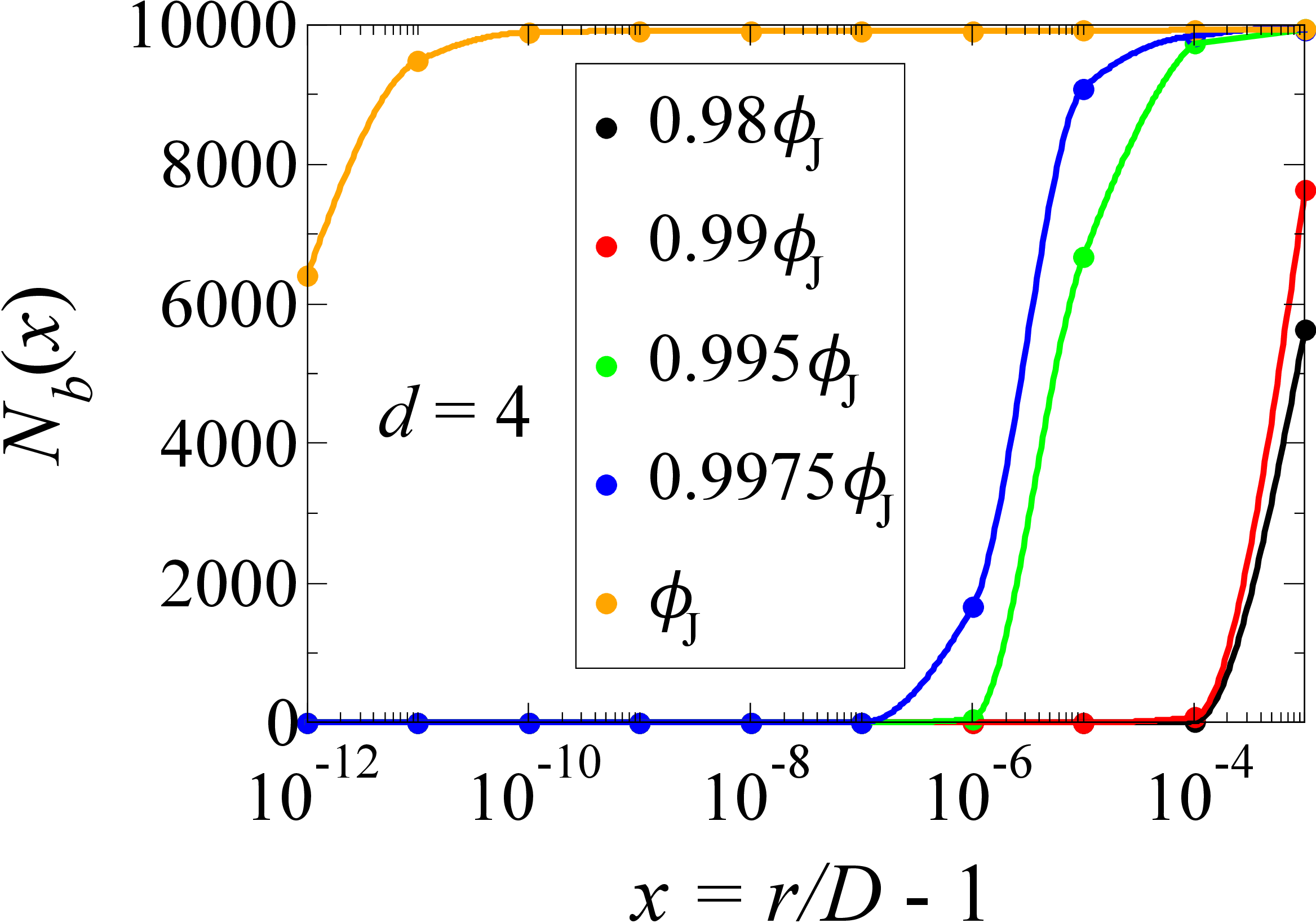}\hspace{0.05\textwidth}}
        \subfloat[]{\includegraphics[height=0.35\textwidth]{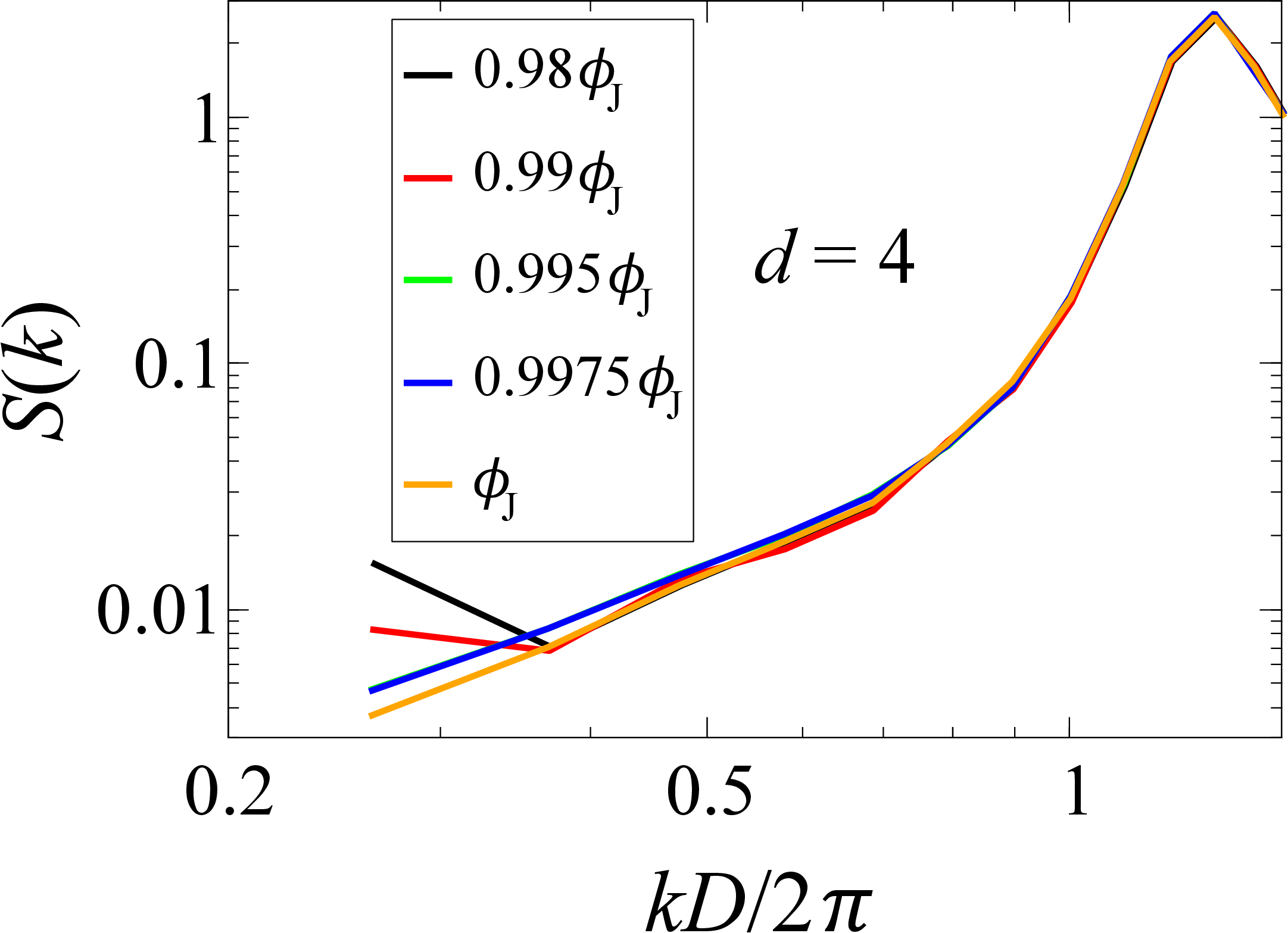}}
        \caption{The (a) number of backbone spheres $N_b(x)$ as a function of the scaled interpaticle gap $r/D - 1$, where $D$ is the particle diameter for several fractions of the jamming packing fraction $\phi_J$=0.452 and (b) the corresponding structure factors $S(k)$ as a function of the scaled wave number $kD/2\pi$ in $d=4$ with $N = 10000$.}
    \end{figure*}

    \begin{figure*}[!t]
        \centering
        \subfloat[]{\hspace{-0.05\textwidth}\includegraphics[height=0.35\textwidth]{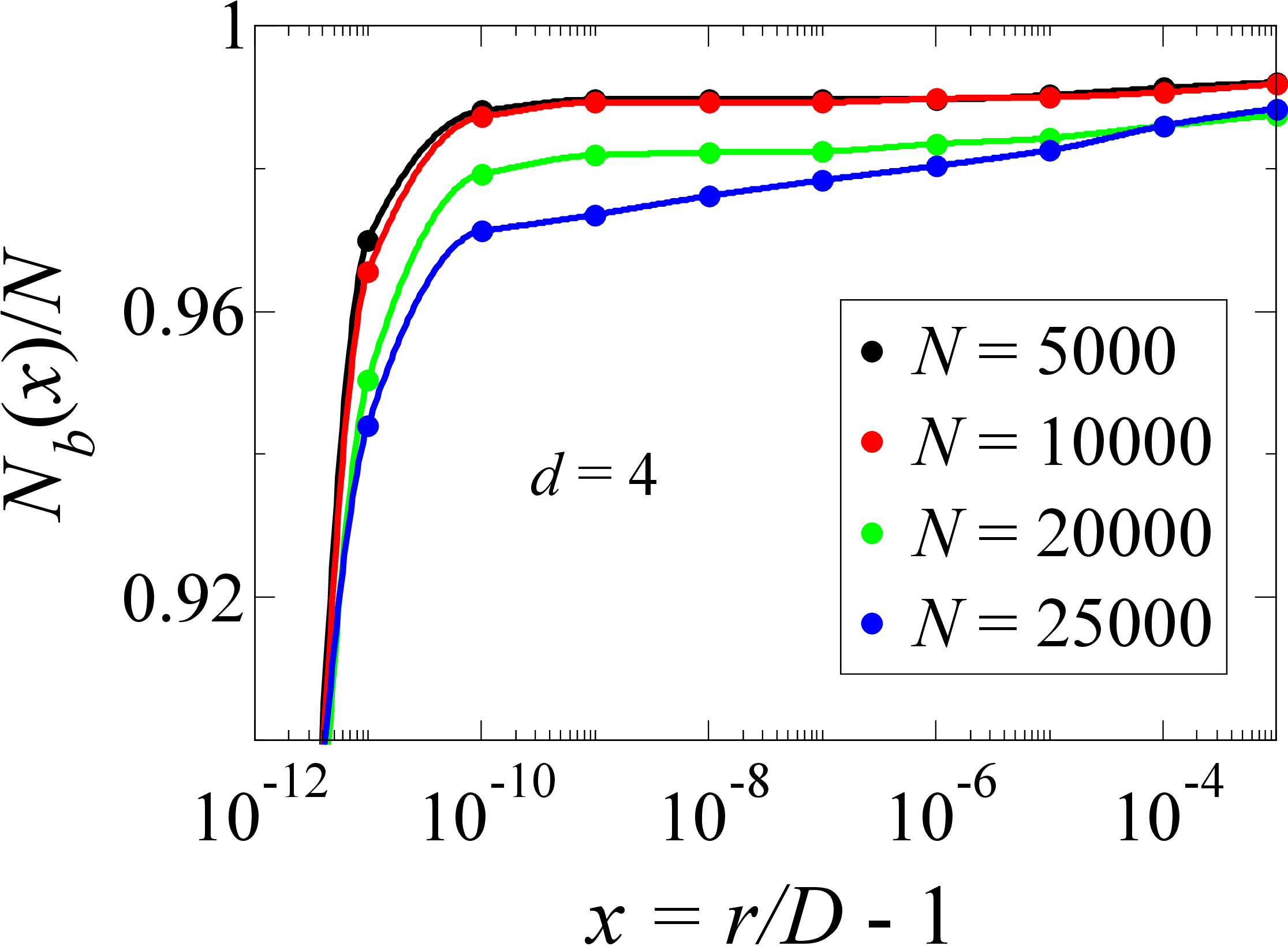}\hspace{0.05\textwidth}}
        \subfloat[]{\includegraphics[height=0.35\textwidth]{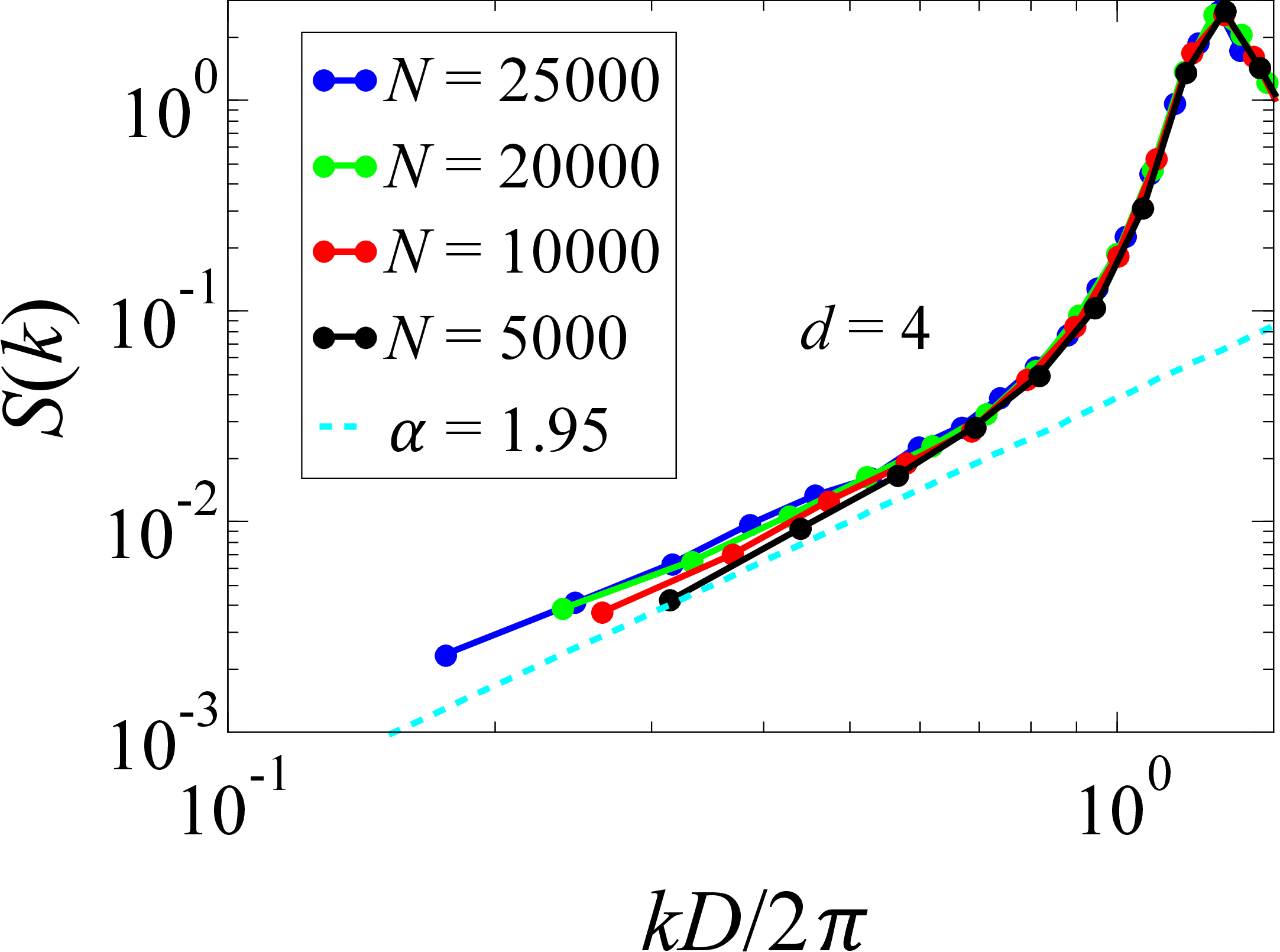}}
        \caption{The number of backbone sphere $N_b(x)$ scaled by the number of spheres as a function of the scaled interparticle gap $r/D - 1$, where $D$ is the particle diameter and (b) the corresponding structure factors $S(k)$ as a function of the scaled wave number $kD/2\pi$ in $d = 4$ for packings with $N = 5000$ ($\phi_J = 0.452$),  $N = 10000$ ($\phi_J = 0.452$),  $N = 20000$ ($\phi_J = 0.453$), and  $N = 25000$ ($\phi_J = 0.453$). (b) has a dashed cyan line indicating $\alpha = 1.95$ scaling.}
    \end{figure*}

\section{Generation and Characterization of 4D and 5D MRJ Hypersphere Packings}
Having established that ensuring strict jamming results in hyperuniform 3D MRJ sphere packings with $\alpha=0.973\pm0.02$, we turn our attention to 4D and 5D hyperspheres to characterize the hyperuniformity of their strictly jammed MRJ states.
In the following subsections, we provide evidence suggesting it is more important to achieve strict jamming \textcolor{black}{to} ensure hyperuniformity in 4D and 5D hypersphere packings than in 3D sphere packings.
To do so, we generate 4D and 5D MRJ hypersphere packings and conduct the same two numerical experiments carried out on the 3D packings.
We then examine the trends in hyperuniformity scaling exponent $\alpha$ and $S(k)$ for packings of 3D, 4D, and 5D (hyper)spheres.

\begin{figure*}[!t]
        \centering
        \subfloat[]{\hspace{-0.05\textwidth}\includegraphics[height=0.24\textwidth]{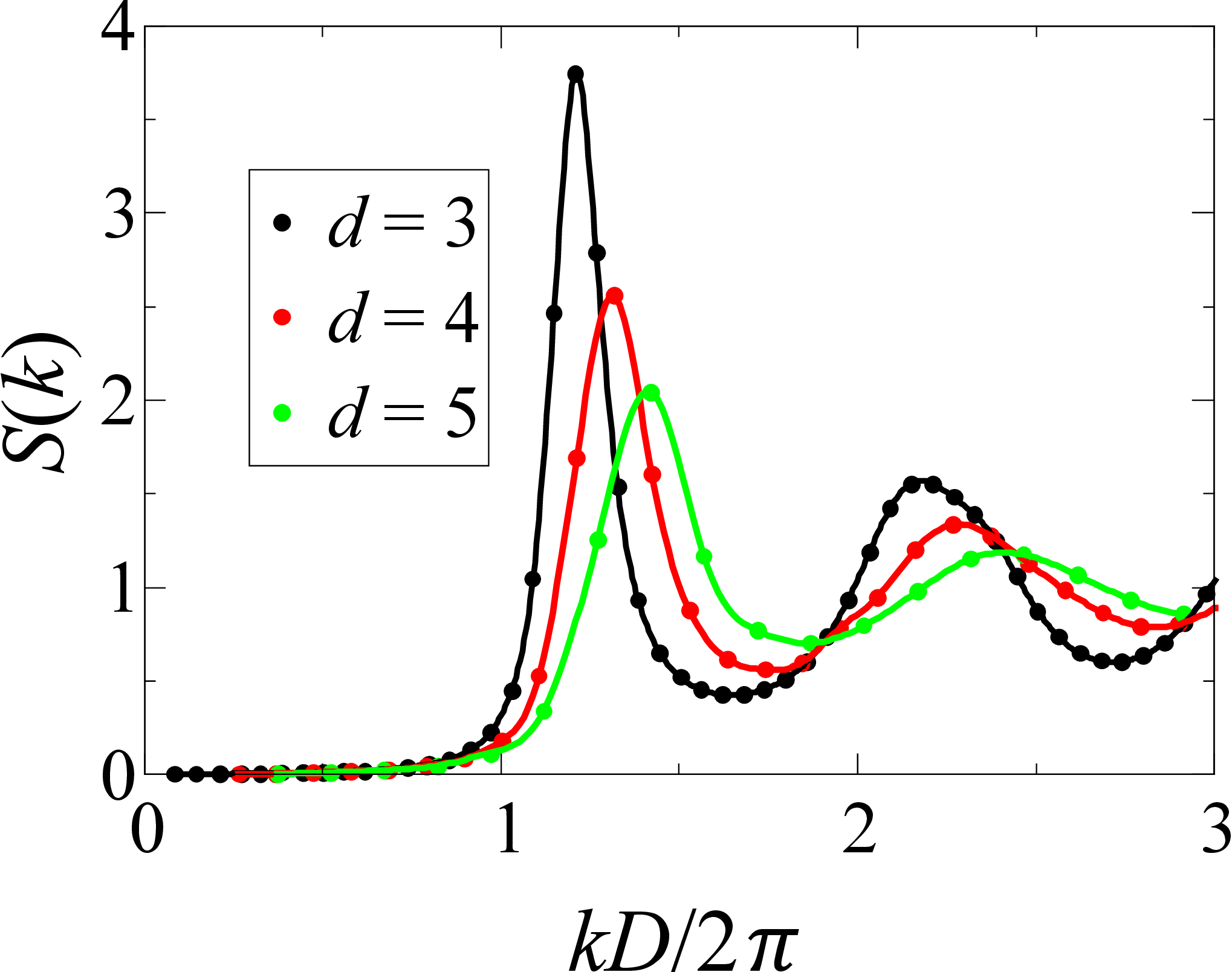}\hspace{0.05\textwidth}}
        \subfloat[]{\includegraphics[height=0.24\textwidth]{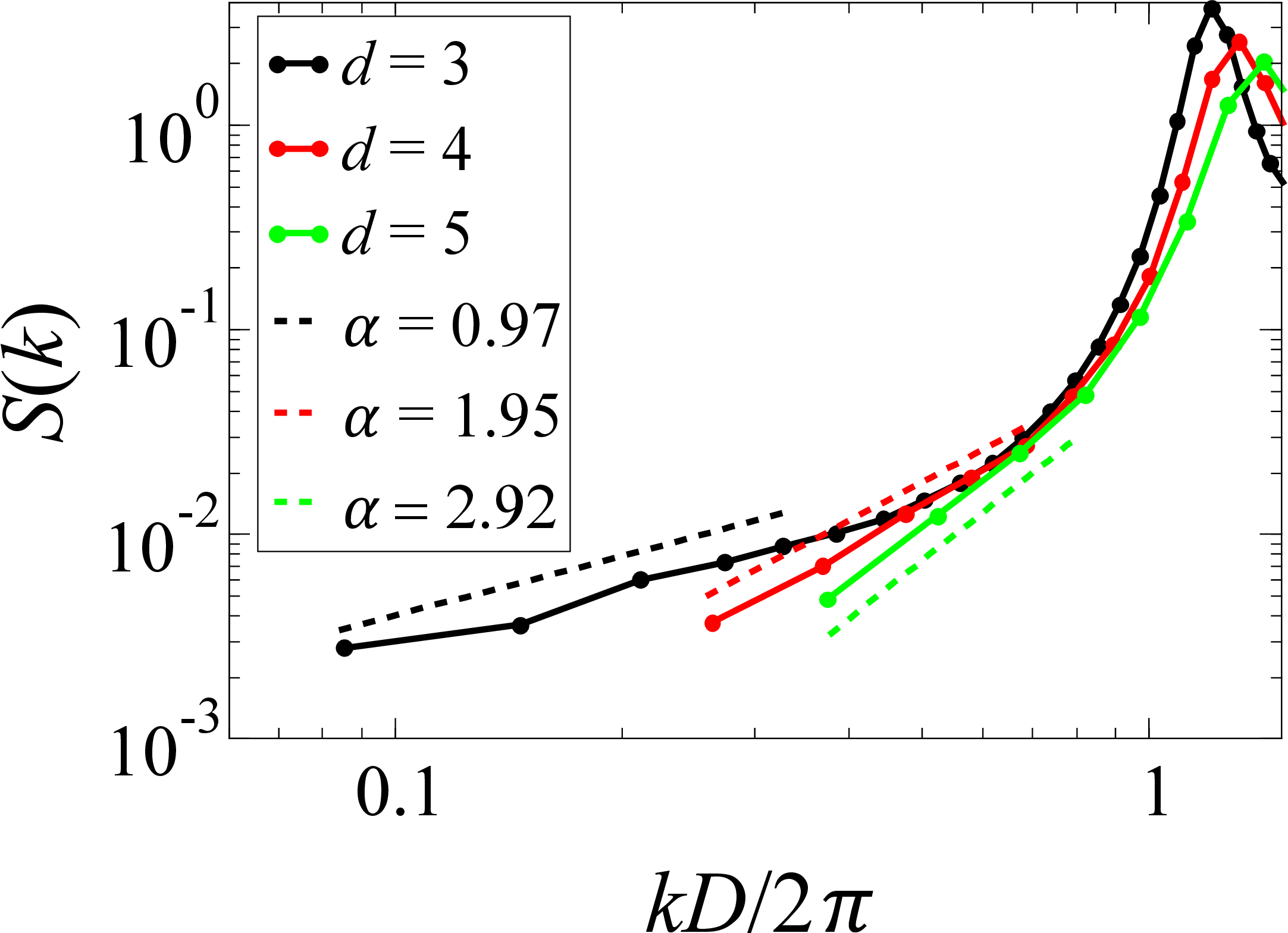}}
        \subfloat[]{\hspace{0.025\textwidth}\includegraphics[height=0.24\textwidth]{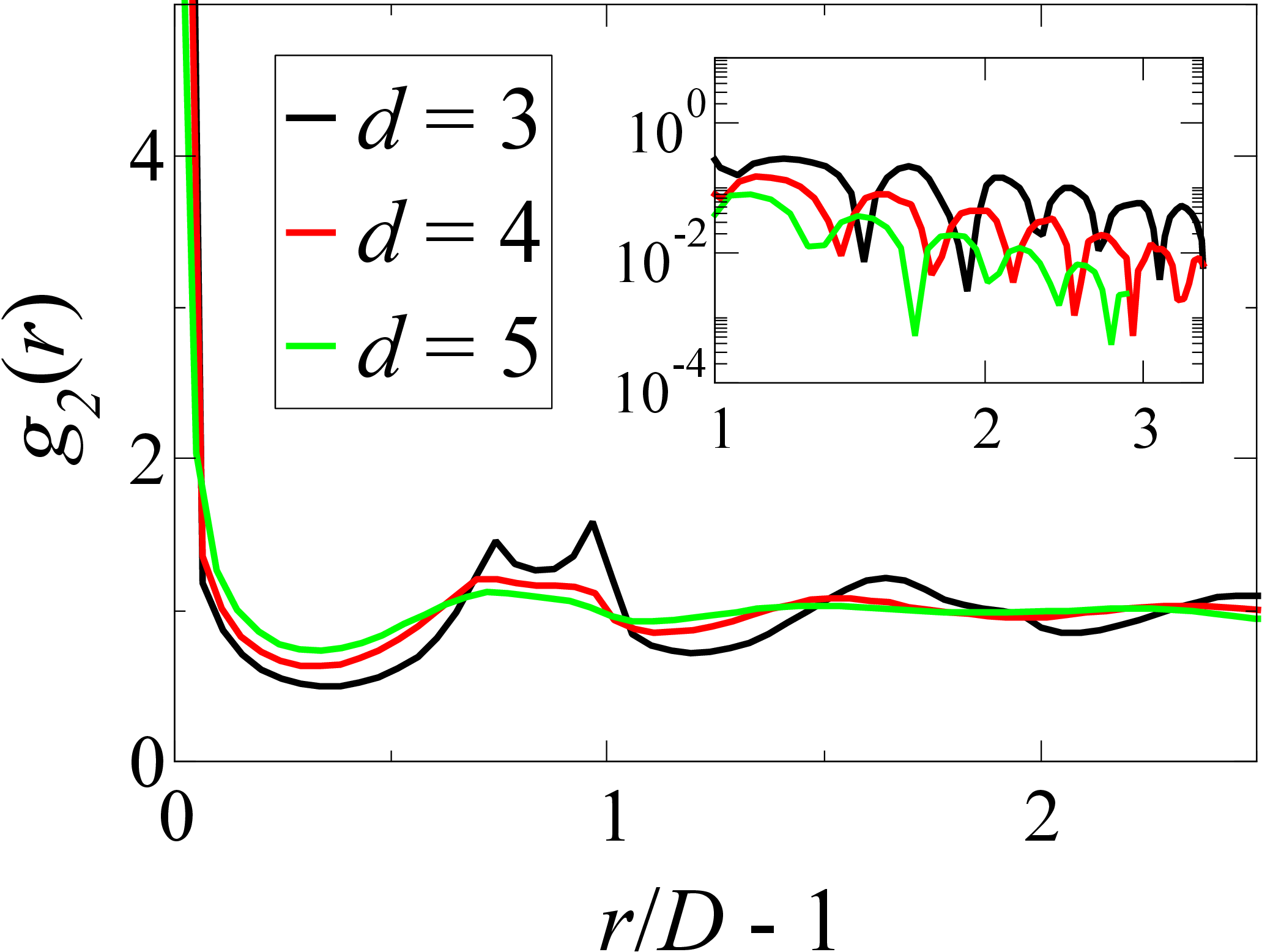}}
        \caption{Ensemble-averaged structure factors $S(k)$ of the MRJ sphere packings in $d = 3,4,5$ (with $N = 5000 \;\textrm{(}\phi_J = 0.639\textrm{)}, 10000 \;\textrm{(}\phi_J = 0.452\textrm{)}, 20000 \;\textrm{(}\phi_J = 0.309\textrm{)}$, respectively) as a function of the scaled wave number $kD/2\pi$, where $D$ is the particle diameter, on a linear (a) and log-log (b) scale.  The black, red, and green dashed lines in (b) show hyperuniformity scaling exponents $\alpha = 0.97, 1.95, 2.8$ respectively. The associated pair correlations $g_2(r)$ are shown in (c), where the inset shows the log scale plot of $|h(r)|$ (i.e., $|g_2(r)-1|$).}
        \label{fig:Sk_all}
    \end{figure*}
    
\subsection{Relationships between hyperuniformity, jamming, and system size are consistent with 3D packings}

Here, we demonstrate that achieving strict jamming in packings of 4D hyperspheres ensures hyperuniformity.
Figure 3 demonstrates that the behaviors of $N_b(x)$ and $S(k)$ as $\phi$ approaches $\phi_J$ for 4D hyperspheres are consistent with \textcolor{black}{those of 3D} spheres. 
By comparing the curves in Figs. 1(a) and 3(a), it is clear that the backbone degrades much more quickly as $\phi$ decreases in 4D packings than in 3D packings.
This is accompanied by a significantly larger increase in the smallest-$k$ value of $S(k)$, indicating a stronger increase of nonhyperuniform character [cf. Figs. 1(b) and 3(b)].
From this, we conclude that ensuring high-quality strict jamming becomes more important as $d$ increases when considering the hyperuniformity of MRJ hypersphere packings.
Figure 4(a) indicates that the largest 4D packing that reliably reaches strict jamming is $N = 10000$. 
Like the 3D packings, above this system size the integrity of the backbone decreases, which is accompanied by an increase in the magnitude of the small-$k$ regime of $S(k)$ [see Fig. 4(b)].
This further supports the notion that larger packings appear less hyperuniform because they cannot be jammed to the same quality as the smaller packings. In a similar fashion, the largest 5D packing that reliably reaches strict jamming is determined to be $N = 20000$.

\subsection{Structure factors and hyperuniformity scaling exponents}

We now compare $S(k)$, for the largest-$N$ ensembles, strictly jammed packings of 3D, 4D, and 5D (hyper)spheres.
Compared to those given by Skoge {\it et al.} \cite{MRJ_HighDim}, we find the \textcolor{black}{principal} peak heights in our $S(k)$ are somewhat smaller.
We attribute this greater local disorder to the TJ step of the hybrid scheme finding the inherent structure via a steepest-descent trajectory \cite{TJ_Alg}, as opposed to a pure LS scheme that allows for dynamical equilibration of the particles and can result in the formation of crystalline domains \cite{Jiao_Diversity}. 
The increased disorder of the hybrid scheme is crucial when trying to produce MRJ-like states.
The attenuation and broadening of the peaks in $S(k)$ in Fig. 5(a) as $d$ increases is consistent with the \textit{decorrelation principle}, which states that \textit{unconstrained} spatial correlations vanish asymptotically for pair distances beyond the particle diameter in the high-dimensional limit and any higher-order correlation functions $g_n(\mathbf{r}_1,\dots,\mathbf{r}_n)$ may be expressed up to small errors in terms of the number density $\rho$ and the pair correlation function $g_2(\mathbf{r})$ \cite{DecorPrinc}.
This implies that as $d\rightarrow\infty$, $g_2$ for an MRJ packing would comprise a radial Dirac delta function at contact $\delta(r=D)$ and a Heaviside step function for all pair distances greater than $D$ $\Theta(r-D)$ \cite{DecorPrinc}. 
Decorrelation is already exhibited in relatively low dimensions, as can be clearly seen in pair correlation functions shown in Fig. \ref{fig:Sk_all}(c), which is consistent with those reported in Ref. \cite{MRJ_HighDim}.

Figure 5(b) shows that $\alpha$ increases roughly by 1 when $d$ increases by 1 (see Table I).
To reiterate, these packings do not represent the \textit{ideal} MRJ state because they contain rattlers, and thus, are less hyperuniform than we expect the ideal MRJ state to be \cite{MRJ_CritSlow,Sal_AB}.
These results suggest that for the ideal MRJ state, $\alpha$ may scale as $\alpha=d-2\;\forall\;d\geq3$. Based on Eq. (\ref{eq:classes}), this also indicates that MRJ packings are class-II hyperuniform for $d=3$, and class-I hyperuniform for all $d \ge 4$ \cite{03HU}.
Note that this implies that in the infinite-dimensional limit, MRJ packings become stealthy-like \cite{Steal1}, i.e., that structure factor becomes perfectly flat and equal to zero in the vicinity of the origin.
Moreover, Fig. 5(a) clearly shows the stealthy-like flattening of $S(k)$ near the origin begins at larger wave numbers as dimension increases.

For $d = 3$, $\alpha = 1$ indicates $S(k)$ is nonanalytic at the origin, implying a power-law large-$r$ asymptotic behavior in the pair-correlation function, i.e., $|h(r)|=|g_2(r)-1| \sim 1/r^4$. 
An analytic $S(k)$ possesses only even powers of $k$ and arises whenever $g_2(r)$ decays exponentially fast or faster \cite{HU_PhysRep}.
While our results suggest $\alpha = 2$ for $d=4$ and might imply analytic behavior, it would be inconsistent to find nonanalytic behavior for $d = 3$ and not for $d = 4$.
Nonanalyticity for $d=4$ with $\alpha = 2$ would require higher odd powers in $k$ in the small-$k$ expansion of $S(k)$.
Indeed, we find by carefully fitting the data that $S(k) \sim c_2 k^2 + c_3 k^3$ at the origin with $c_3/c_2 \approx 1/20$, which indicates $S(k)$ is nonanalytic at the origin. 
Moreover, nonanalytic behavior is consistent with observed power-law decay of $g_2(r)$ [see Fig. \ref{fig:Sk_all}(c)]. 
The non-analyticity of $S(k)$ for $d=5$, which has $\alpha = 3$, also leads to observed power-law decay in the corresponding pair correlations, i.e., $|h(r)| \sim 1/r^6$.

Additionally, we find that the rattler fraction $\phi_R$ in the numerically generated MRJ configurations decreases rapidly as $d$ increases (cf. Table I), consistent with previous findings \cite{TJ_2D, TJ_Alg, MRJ_HighDim}, and expect it to vanish for sufficiently large $d$. 
The decrease of $\phi_R$ is associated with an exponentially increasing kissing number (i.e., the number of spheres that can touch a central sphere without overlapping), and implies the packings become more uniform locally as $d$ increases, which collectively leads to an improved degree of hyperuniformity on the global scale. 
    
\begin{table}[!h]
\centering
\caption{The rattler fractions $\phi_R$ and hyperuniformity scaling exponents $\alpha$ extracted from the structure factors of the 3D, 4D, and 5D MRJ (hyper)sphere packings using the long-time scaling of the excess spreadability. These values agree with the direct-fit values within error. \textcolor{black}{The uncertainty in these $\alpha$ measurements is the standard deviation of the $\alpha$ values extracted from the excess spreadabilities for each packing individually.}}
\begin{tabular}{|l|l|l|l|}
\hline 
$d$ & $N$   & $\phi_R$ & $\alpha$        \\\hline 
3   & 5000  & 1.6\% & 0.973$\pm$0.020 \\\hline 
4   & 10000 & 0.96\% & 1.946$\pm$0.013 \\\hline 
5   & 20000 & 0.57\% & $2.926\pm0.012$ \\\hline 
\end{tabular}
\end{table}

\section{Discussion}
We used a hybrid LS-to-TJ scheme to generate large strictly jammed MRJ packings of identical hyperspheres in dimensions three through five.
We conducted two different numerical experiments on these packings.
First, for a fixed number of spheres $N$, we tracked how the quality of the jammed backbone and small-$k$ behavior of $S(\mathbf{k})$ behaved as the strict jamming point was approached.
Then, we examined how these same two structural descriptors varied as $N$ increased.
In each spatial dimension, we found that as the strict jamming point was approached, the quality of the backbone increased sharply with a concomitant decrease in the magnitude of the small-$k$ region of $S(k)$, signaling an increase in the degree of hyperuniformity of the packings.
Moreover, as $N$ increased, the quality of the backbone decreased,while the magnitude of the small-$k$ region of $S(\mathbf{k})$ increased, indicating the onset of nonhyperuniformity due to strict jamming becoming more difficult to achieve.
These results vividly demonstrated that achieving strict jamming is critical to ensuring hyperuniformity and extracting the precise value of the hyperuniformity scaling exponent $\alpha$. 

We extracted $\alpha$ from the largest strictly jammed packings we produced by fitting the long-time behavior of the excess spreadability \cite{Spread_Origin} and found $\alpha = 0.973\pm0.20, 1.946\pm0.13,$ $2.926\pm0.12$ in $d=3,4,5$, respectively.
In addition, we found that $\phi_R$ decreased substantially as $d$ increased, implying that the ideal MRJ state is easier to approach as $d$ increases.
Note that this value of $\alpha$ in $d=3$ is markedly larger than the value reported by Wilken et al. in Ref. [\citenum{BRO_origin}] ($\alpha = 0.25$) for Manna-class RCP BRO packings and from the LS-based packings produced by Donev {\it et al.} \cite{Donev_1m} ($\alpha\approx0.46$, see Fig. 6).
Moreover, the larger positive values of $\alpha$ for the 4D and 5D MRJ packings, indicating stronger hyperuniformity, are diametrically opposite to the \textit{nonhyperuniformity} ($\alpha = 0$) of the 4D and 5D Manna-class RCP BRO packings reported by Wilken {\it et al.} in Ref. [\citenum{BRO_highdim}].
These stark differences in $\alpha$ in dimension $d = 3,4,$ and 5 vividly illustrate the distinctions between MRJ and RCP states, which was previously shown for the corresponding 2D states \cite{TJ_2D}.
In the latter case, Ref. [\citenum{TJ_2D}] suggested the existence of 2D \textcolor{black}{monodisperse} isostatic MRJ states with packing fraction $\phi\approx0.83$, which is appreciably less than that of hyperstatic RCP packings, which are identified as the most probable packings \textcolor{black}{or endpoints of certain dynamical algorithms \cite{BRO_origin}}, a physically meaningful definition of RCP \cite{Isostat_Ohern}.

\begin{figure*}[t!]
    \centering
    \subfloat[]{\includegraphics[height=0.33\textwidth]{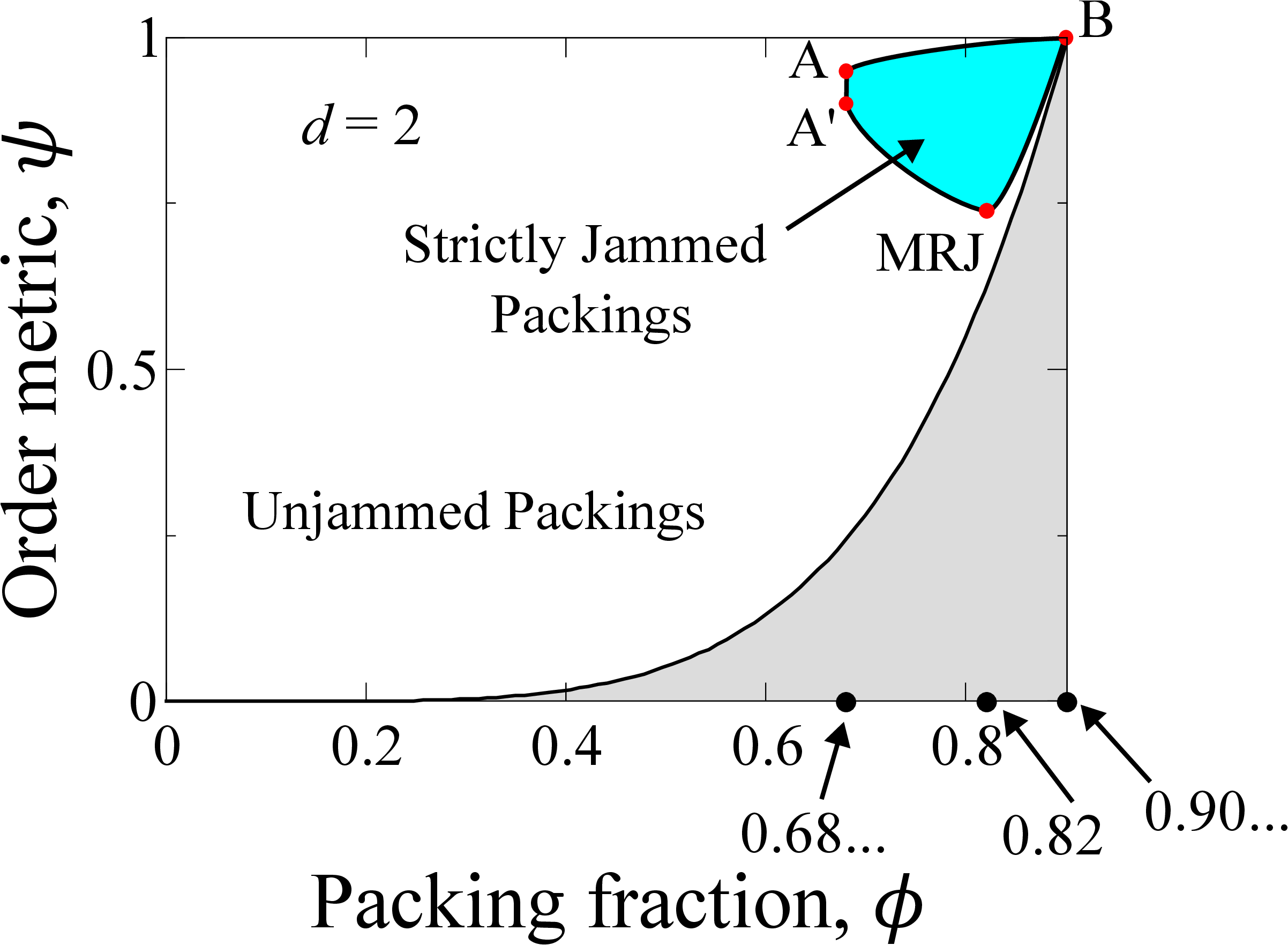}}
    \subfloat[]{\includegraphics[height=0.33\textwidth]{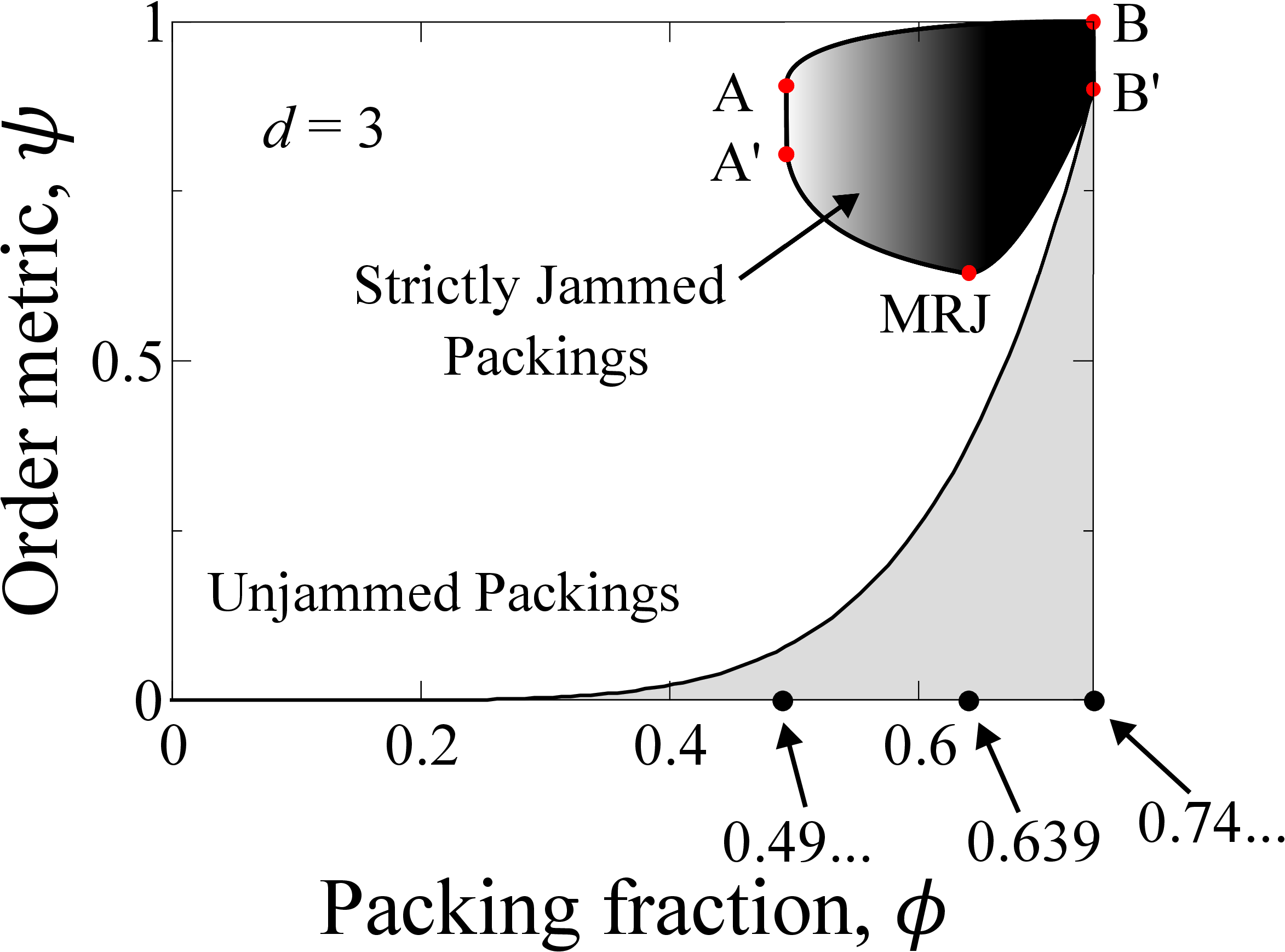}}\\
    \subfloat[]{\includegraphics[height=0.33\textwidth]{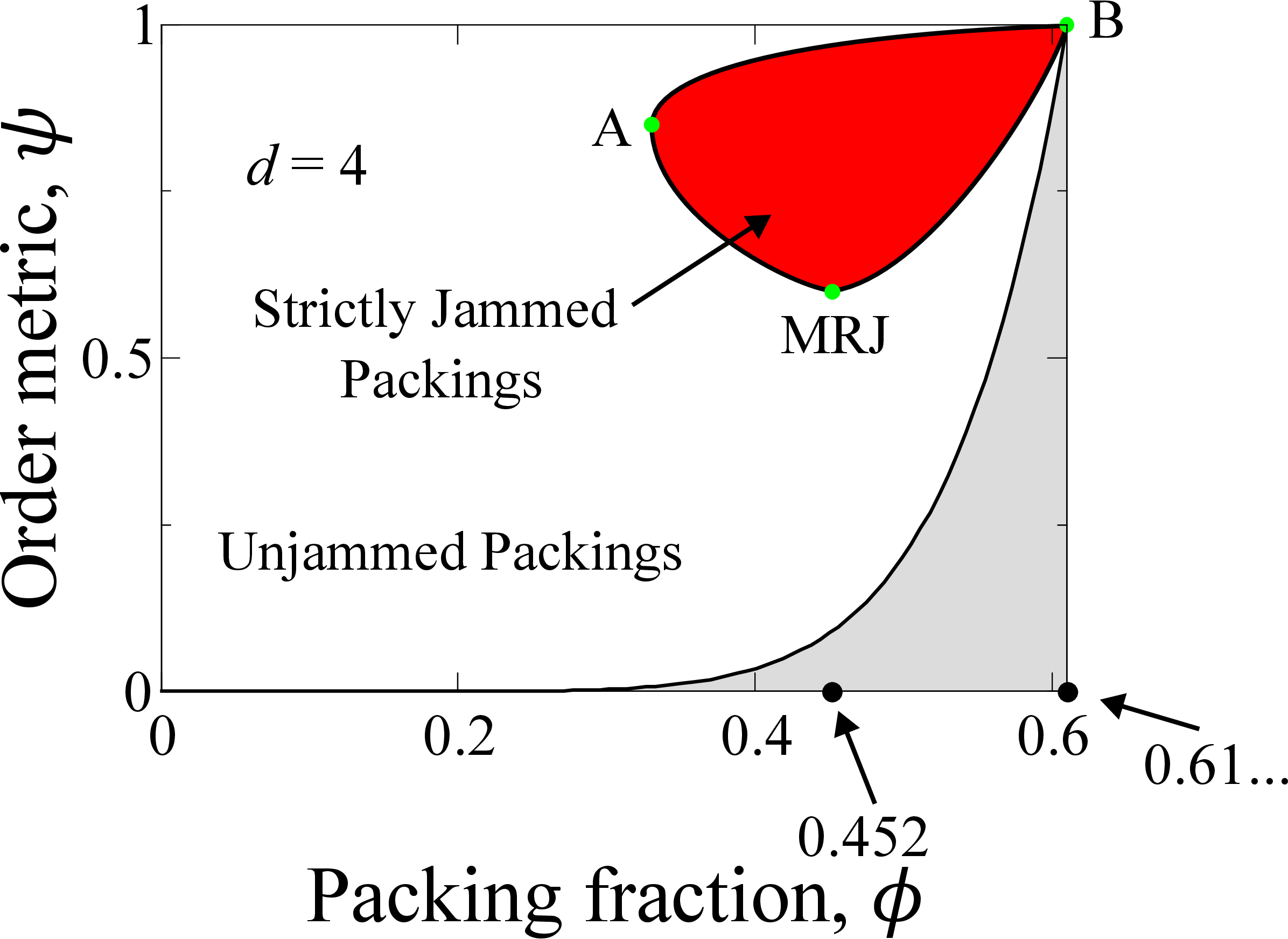}}
    \subfloat[]{\includegraphics[height=0.33\textwidth]{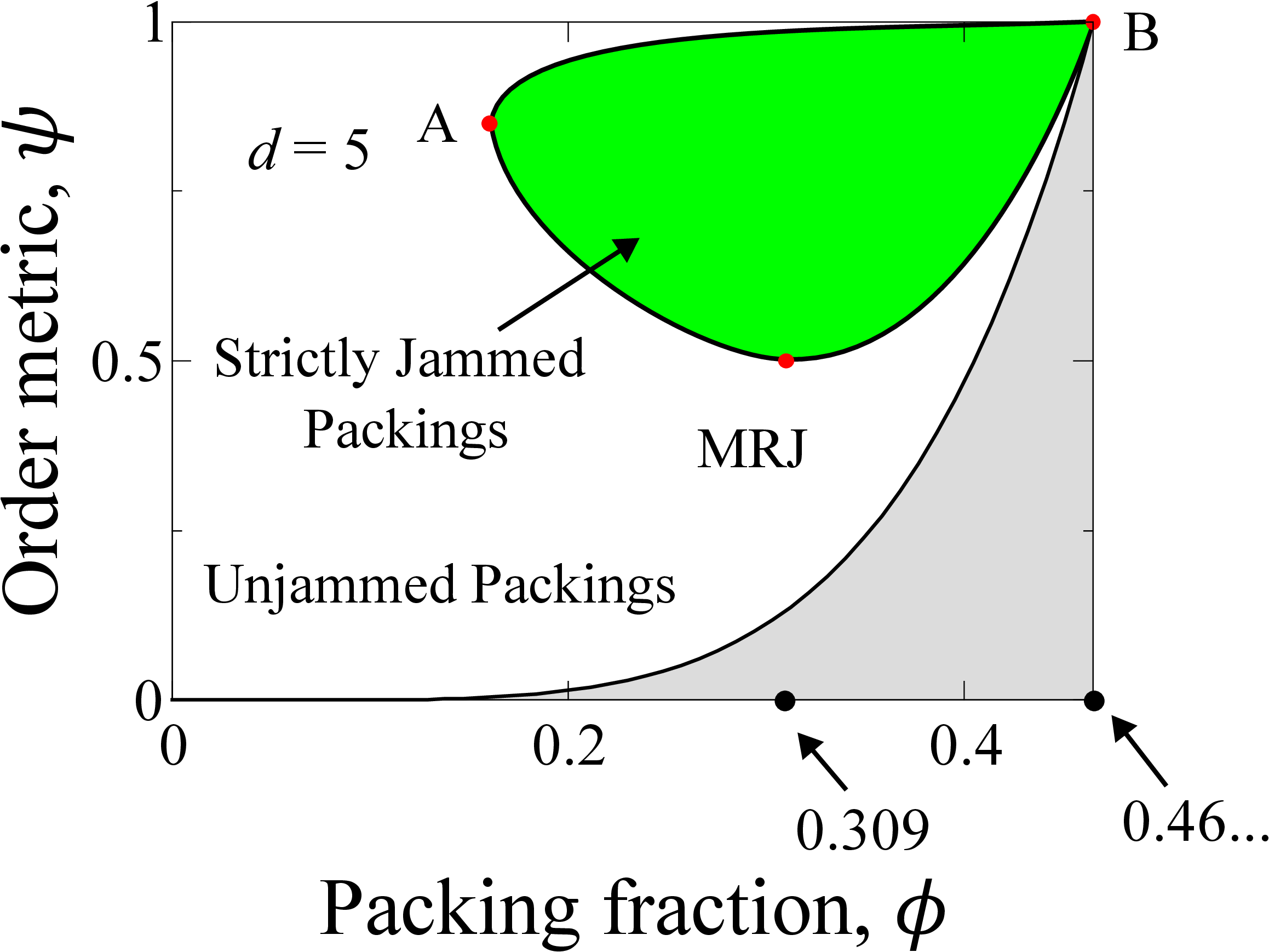}}\\
    \caption{Schematic figure adapted from Refs. \citenum{PackingRev} and \citenum{TJ_2D} depicting order maps represented in the $\phi-\psi$ plane for (a) 2D, (b) 3D, (c) 4D, and (d) 5D strictly jammed, frictionless, monodisperse hard-hypersphere packings. The cyan (a), black gradient (b), red (c), and green (d) regions depict the space of strictly jammed packings, where the MRJ point, representing the maximally random jammed state, can be seen at the bottom of each of these regions. The white regions correspond to unjammed packings, and the gray region corresponds to no packings. In (a), the locus of points A$-$A' corresponds to the lowest-density jammed packings, conjectured to be the reinforced kagom{\'e} lattice, reinforced rectangular kagom{\'e} lattice, and other combinations all with $\phi_{\textrm{min}} = 0.6801...$ \cite{PTest_1}, the point B corresponds to the triangular lattice with $\phi_{\textrm{max}} = 0.9068...$, and the MRJ state has $\phi_{\textrm{MRJ}} = 0.82$. In (b), the locus of points A$-$A' corresponds to the lowest-density jammed packings, conjectured to be tunneled crystals with $\phi_{\textrm{min}} = 0.4936...$ \cite{tunnel}, the locus of points B$-$B' corresponds to the fcc sphere packing and its stacking variants all with $\phi_{\textrm{max}} =0.7404...$, and the MRJ state has $\phi_{\textrm{MRJ}} = 0.639$. In (c), A corresponds to the presently unknown lowest-density jammed state, B corresponds to the checkerboard-lattice packing $D_4$ with $\phi_{\textrm{max}} = 0.6168...$ (densest-known packing for $d=4$) \cite{HighDimCryst}, and the MRJ state has $\phi_{\textrm{MRJ}} = 0.452$. In (d), A corresponds to the presently unknown lowest-density jammed state, B corresponds to the checkerboard-lattice packing $D_5$ with $\phi_{\textrm{max}} = 0.4652...$ (densest-known packing for $d=5$) \cite{HighDimCryst}, and the MRJ state has $\phi_{\textrm{MRJ}} = 0.309$.}
    \label{fig:Maps}
\end{figure*}

\begin{figure}[t!]
\centering
\includegraphics[width=0.45\textwidth]{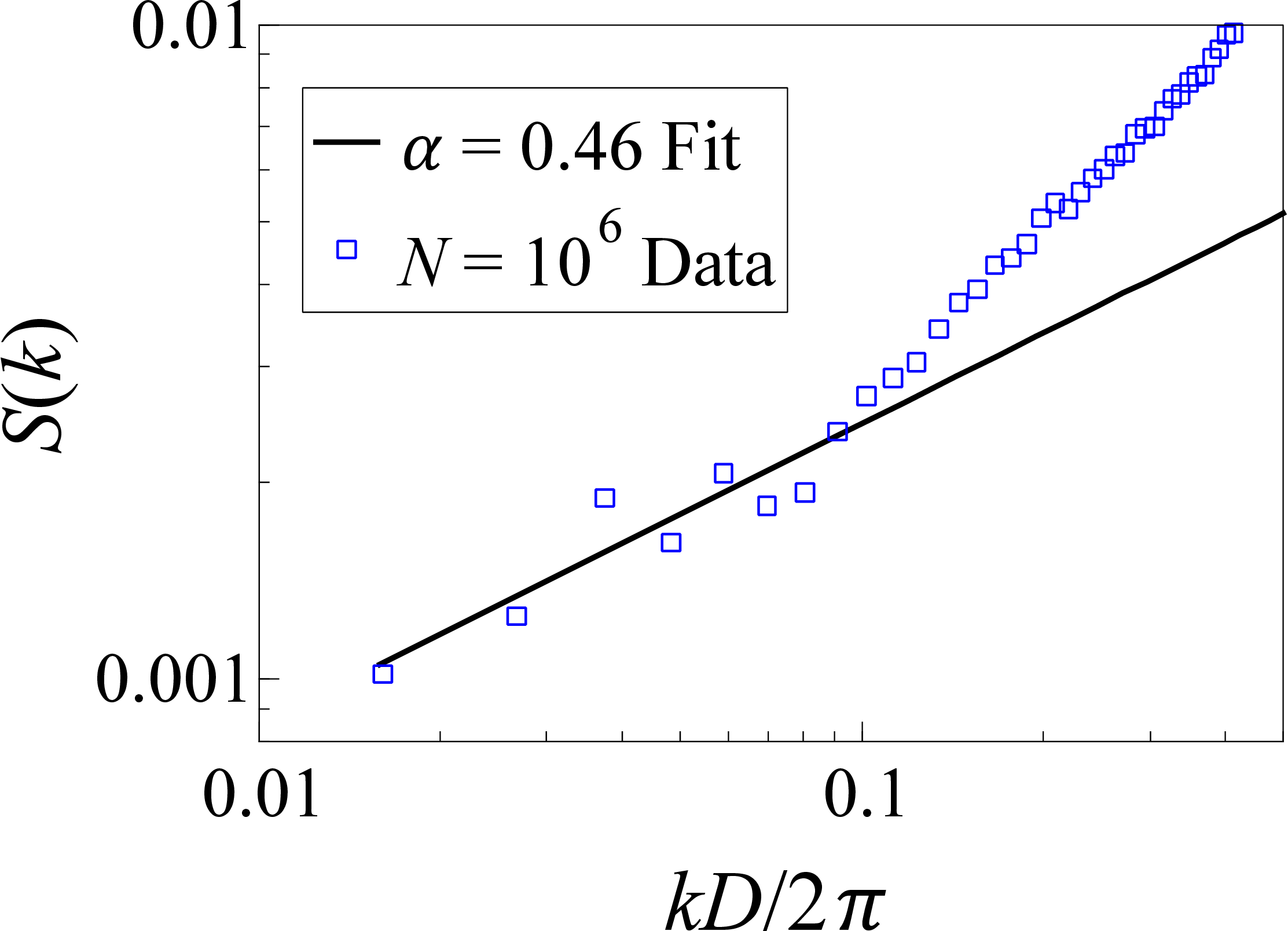}
\caption{Structure factor $S(k)$ for a $N = 10^6$ sphere packing as a function of the scaled wave number $kD/2\pi$, where $D$ is the particle diameter, reproduced from Ref. [\citenum{Donev_1m}] (blue squares). The solid black line is a fit to the small-$k$ region ($k \lesssim 0.1$) with a power-law scaling of $\alpha = 0.46$.}
\label{fig:Donev}
\end{figure}

\begin{figure}[h!]
    \centering
    \subfloat[]{\includegraphics[width=0.45\textwidth]{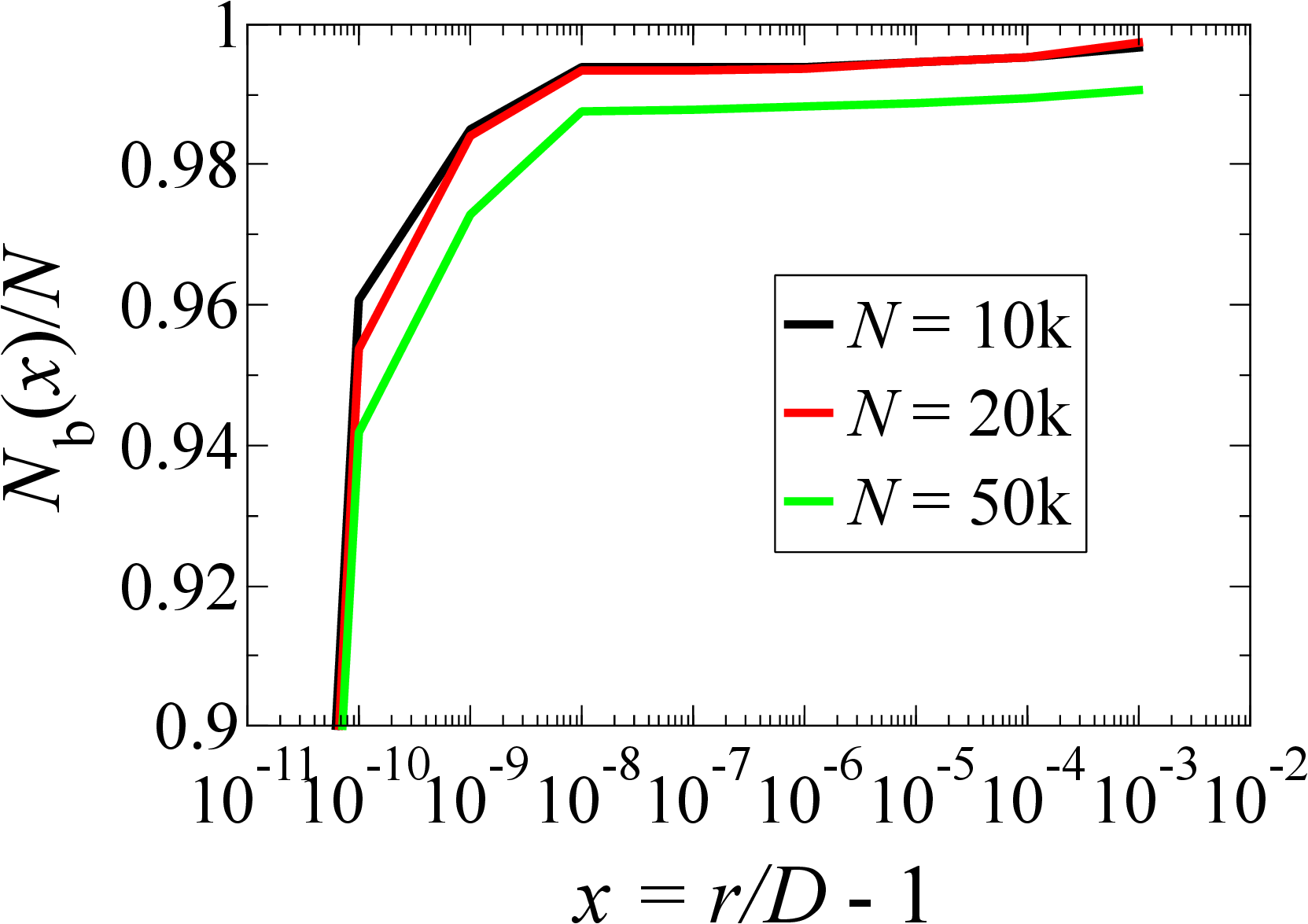}}\\
    \subfloat[]{\includegraphics[width=0.45\textwidth]{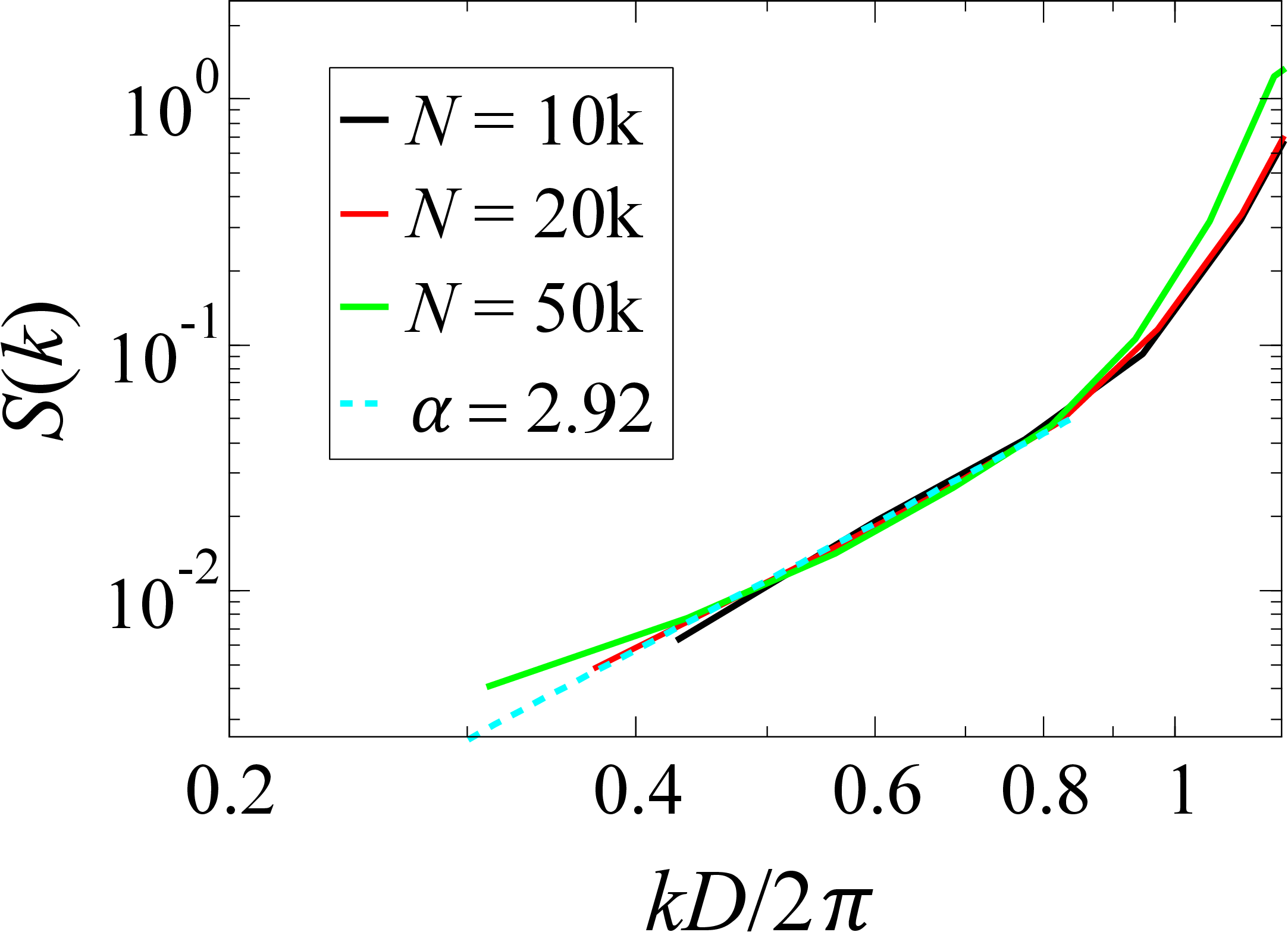}}
    \caption{The number of backbone spheres $N_b(x)$ scaled by the number of spheres as a function of the scaled interparticle gap $r/D - 1$, where $D$ is the particle diameter and (b) the corresponding structure factors $S(k)$ as a function of the scaled wave number $kD/2\pi$ in $d = 5$ for packings with $N = 10000, 20000, 50000$. (b) has a dashed cyan line indicating $\alpha = 2.92$ scaling.}
    \label{fig:5D}
\end{figure}

\textcolor{black}{Importantly, MRJ states are defined by a geometric-structure approach, which emphasizes the analysis of {\it individual} packings, regardless of the probability of their appearance \cite{PackingRev, MRJ_defn, PTest_1}. Among jammed configurations, the MRJ state is the most disordered configuration (in the infinite-volume limit), as measured by a set of sensitive scalar order metrics, subject to a particular jamming category, which in this case is strict jamming and hence isostatic \cite{PackingRev, MRJ_defn, PTest_1}. The geometric-structure analysis leads to an {\it order map} for strictly jammed states, schematically shown in Fig. \ref{fig:Maps} for dimensions two through five, in which the MRJ states are the minima. A sensitive order metric $\psi$ has the following general properties: (1) it is a well-defined scalar function of a particle configuration $\mathbf{R}$; (2) it is subject to the normalization $0\leq\psi\leq1$, where $\psi=0$ and $\psi=1$ represent the most disordered and ordered possible structures; and (3) for any two configurations $\mathbf{R_A}$ and $\mathbf{R_B}$, $\psi(\mathbf{R_A}) < \psi(\mathbf{R_B})$ implies $\mathbf{R_A}$ is more disordered than $\mathbf{R_B}$ (see Refs. \cite{PackingRev, PTest_1} for specific examples of sensitive order metrics and additional details). Thus, the MRJ state is clearly distinguished from RCP states, which have been defined as either the most probable jammed configurations within an ensemble  \cite{Isostat_Ohern} or the endpoint of certain dynamical algorithms \cite{BRO_origin}.}

As noted earlier, free-volume theory arguments \cite{Sal_AB} strongly indicate that the \textit{ideal} hyperuniform MRJ state is rattler free, and thus rattlers act as defects in putative MRJ packings.
We surmise that the presence of a small concentration of rattlers results in a decrease in $\alpha$.
Thus, our numerical results suggest the scaling relation $\alpha = d-2$.
\textcolor{black}{The results in Ref. \cite{Zachary_QLRlet} present numerical evidence suggesting $\alpha = 1$ for binary MRJ disk packings. If one assumes that the MRJ monodisperse disk packings numerically generated in Ref. [\citenum{TJ_2D}] are in the same universality class, then one would conclude that $\alpha = 1$ in $d=2$ and $d=3$. Thus, the scaling argument presented here is only applicable for $d \geq 3$, implying} that \textcolor{black}{ideal} 3D MRJ sphere packings are hyperuniform of class II and 4D and 5D hypersphere packings are class-I hyperuniform.
Moreover, \textcolor{black}{the exponent} $\alpha=d-2$ implies that in the infinite-dimensional limit, MRJ packings become stealthy-like \cite{Steal1}.
These findings motivate the future development of an algorithm that produces rattler-free 3D packings of identical particles to approach the ideal MRJ state.

~
\section*{Acknowledgements}{The authors thank H. Wang, M. Skolnick, S. Wilken, P. Chaikin, P. Morse, and S. Chen for insightful discussions. This research was sponsored by the Army Research Office and was accomplished under Cooperative Agreement No. W911NF-22-2-0103 as well as the National Science Foundation under Award No. CBET-2133179.}

\section*{Appendix A: Hyperuniformity of the $N = 10^6$ packings from Ref. [37]}

Here, we reproduce the direct Fourier transform data used to produce the blue squares in the top-left inset of Fig. 2 in Ref. [\citenum{Donev_1m}].
The packings therein are produced using an event-driven molecular dynamics algorithm, which is known to struggle to produce isostatic packings with $N \gtrsim 1000$ \cite{MRJ_PairStat}.
Thus, these packings with $N = 10^6$ spheres are not jammed.
In Ref. [\citenum{Donev_1m}], the authors fit $S(k)$ on a linear scale and conclude that $\alpha = 1$ for these packings.
In Fig. \ref{fig:Donev}, we fit these data on a logarithmic scale and find that $\alpha=0.46$, which is substantially lower than the previously reported value of 1.
The smaller $\alpha$ value observed for these packings is consistent with a 3D sphere packing that is away from jamming.

\section*{Appendix B: System size study of 5D hyperspheres}

Figure \ref{fig:5D}(a) shows the number of backbone spheres as a function of of the interparticle distance for the putatively jammed  packings of 5D hyperspheres with $N = 10000, 20000, 50000$. 
Both the $N = 10000$ packings and $N = 20000$ packings have backbones of similar quality, while there is a distinct drop-off in the the backbone quality of the $N = 50000$ packings.
This is mirrored in Fig. \ref{fig:5D}(b) where the structure factor for the $N = 50000$ packings shows the onset of nonhyperuniform scaling, while the $N = 10000$ and 20000 packings clearly indicate hyperuniform scaling.
We choose to examine the $N = 20000$ packings here since they are the largest packings we were able to generate in $d = 5$ with a well-defined jammed backbone.

\providecommand{\noopsort}[1]{}\providecommand{\singleletter}[1]{#1}%

\end{document}